%% file: eff_pot.tex
\documentclass[a4paper,12pt]{article}

\usepackage[english]{babel}
\usepackage{graphicx,epsfig,bbm}

\usepackage{cite}
\usepackage{mathrsfs,amsmath}
\usepackage{amssymb}
\usepackage{slashed}

\newcommand{\be}{\begin{equation}}
\newcommand{\ee}{\end{equation}}
\newcommand{\bea}{\begin{eqnarray}}
\newcommand{\eea}{\end{eqnarray}}
\newcommand{\ba}{\begin{array}}
\newcommand{\ea}{\end{array}}
\newcommand{\balg}{\begin{align}}
\newcommand{\ealg}{\end{align}}
\newcommand{\nn}{\nonumber}

\newcommand{\trm}[1]{\textrm{#1}}

\newcommand{\mcl}[1]{\mathcal{#1}}
\newcommand{\mbb}[1]{\mathbb{#1}}
\newcommand{\msc}[1]{\mathscr{#1}}

\newcommand{\ol}[1]{\overline{#1}}

\newcommand{\KK}{\mathcal{K}}
\newcommand{\NK}{\widetilde{N}}
\newcommand{\mK}{\widetilde{m}}
\newcommand{\one}{\mathbb{1}}

\begin{document}

\begin{titlepage}
\vspace*{-2cm}
\flushright{FTUAM 09-30\\ IFT-UAM/CSIC 09-52\\DFPD-09/TH/21}
\vskip 1.5cm
\begin{center}
{\Large\bf One-loop effective potential in $\mcl{M}_4\times T^2$ with and 

\vspace{.2cm}
without 't Hooft flux
\vspace{3mm}}
\end{center}
\vskip 0.5cm
\begin{center}
{\large A.F.~Faedo},$\,^a$ 
{\large D.~Hern\'andez},$\,^b$ 
{\large S.~Rigolin}$\,^c$ 
{and \large M.~Salvatori}$\,^d$
\vskip .1cm
\vskip .7cm
$^a\,$ INFN Sezione di Padova,\\
Via Marzolo 8, I-35131 Padova, Italy  \\
\vskip .1cm
$^b\,$  Departamento de F\'{\i}sica Te\'{o}rica  and Instituto de F\'{\i}sica Te\'{o}rica, \\ 
Universidad Aut\'{o}noma de Madrid,  Cantoblanco, E-28049 Madrid, Spain \\
\vskip .1cm
$^c\,$ Dipartimento di Fisica ``Galileo Galilei'', Universit\`a degli studi di Padova and INFN Padova, Via Marzolo 8, I-35131 Padova, Italy \\
$^d\,$ Altacontrol SW, Madrid, Spain
\vskip .1cm
\end{center}
\vskip 0.5cm

\begin{abstract}
We review the basic notions of compactification in the presence of a background flux. In 
extra-dimentional models with more than five dimensions, Scherk and Schwarz boundary 
conditions have to satisfy 't Hooft consistency conditions. Different vacuum configurations 
can be obtained, depending whether trivial or non-trivial 't Hooft flux is considered. The presence 
of the ``magnetic'' background flux provide, in addition, a mechanism for producing four-dimensional 
chiral fermions.
Particularizing to the six-dimensional case, we calculate the one-loop effective potential for a 
$U(N)$ gauge theory on $\mathcal{M}_4 \times \mathcal{T}^2$. We firstly review the well 
known results of the trivial 't Hooft flux case, where one-loop contributions produce the usual 
Hosotani dynamical symmetry breaking. Finally we applied our result for describing, for the first time, 
the one-loop contributions in the non-trivial 't Hooft flux case.
\end{abstract}
\end{titlepage}
\setcounter{footnote}{0}
\vskip2truecm

\newpage

%
\input{intro.tex}
\input{parte_gen.tex}

%
%
\input{fermions.tex}
%
%
%
%
\input{one_loop_new.tex}
%
%
\input{pheno_new.tex}
%
%
\input{conclu.tex}

%
%
\section*{Acknowledgments}
We are indebted for very useful discussions to E. Alvarez, M.B. Gavela and M. Garc\'{\i}a-P\'erez. 
The work of A.F. Faedo, D. Hernandez and S. Rigolin has been partially supported by CICYT through the 
project FPA2006-05423 and by CAM through the project HEPHACOS, P-ESP-00346. D. Hern\'andez acknowledges financial support from the MEC through FPU grant AP-2005-3603. S.Rigolin aknowledges also the partial support of an Excellence Grant of Fondazione Cariparo and of the European Programme ``Unification in the LHC era'' under the contract PITN-GA-2009-237920 (UNILHC) . 
%
%
\appendix
%
\input{appendix_new.tex}

%
\input{appendix_heat.tex}

%
%

\input{bib_new.tex}
%
%
\end{document}

%% file: intro.tex
%
\section{Introduction}
%

There is still one sector completely unknown in the Standard Model (SM) of electroweak 
interactions: the Higgs sector. The Higgs boson must exist, either as an elementary 
particle or as a composite resonance. 

In the SM, the Higgs boson is a scalar particle with the appropriate bilinear and 
quadrilinear self-interactions to drive the $SU(2)_{EW} \times U(1)_Y$ spontaneous 
symmetry breaking. All experimental available data agree in indicating that the mass 
of such a state should be of the order of the electroweak scale~\cite{lep}, 
$v\sim {\mathcal{O}}(200)$ GeV. However, in the SM, the Higgs mass parameter is not 
protected by any symmetry and thus can, in principle, get corrections which are 
quadratically dependent on possible higher scales to which the Higgs boson is sensitive.  
Ultimately, the Higgs mass should be sensitive to the scale at which quantum gravity effects 
appear: the Planck scale, $M_{Pl}$. Therefore, from the SM point of view, a Higgs mass at 
the electroweak scale appears ``unnatural''. This represents the essence of the SM hierarchy 
problem. 


Three different mechanisms have been devised in order to eliminate the quadratic sensitivity 
of the Higgs mass to the cutoff scale. In the framework of {\it Supersymmetry}, bosonic and 
fermionic contributions to the quadratic divergences cancel each other in such a way that the 
Higgs mass remains affected only by a logarithmic sensitivity to the cutoff scale. In models like 
{\it Technicolor} \cite{technicolour} and {\it Little Higgs} \cite{littlehiggs} the Higgs is a Goldstone 
boson of a global custodial symmetry that is only softly broken. In the last mechanism, 
{\it Gauge-Higgs Unification} \cite{Manton}, the Higgs is a component of a higher dimensional 
gauge multiplet. Its lightness is guaranteed by the gauge symmetry itself.
Independently of the precise nature assumed for the Higgs field, all these proposals require, 
in one way or another, the appearance of new physics at about the TeV scale. The first two 
approaches have been, and are being, intensely studied. However, they tend to be afflicted by  
rather severe fine-tuning requirements (see for example \cite{Casas} for a comprehensive 
review for Supersymmetry and Little Higgs models) when confronted with present experimental 
data. Here, instead, we follow the last and less explored possibility: Gauge-Higgs unification. 

The main idea of Gauge-Higgs unification is that a single higher dimensional gauge field 
gives rise to all the four-dimensional ($4D$) bosonic degrees of freedom: the gauge bosons, 
from the ordinary four space-time components and the scalar bosons (and the Higgs fields among 
them) from the extra-dimensional ones. The essential point concerning the hierarchy problem 
solution is that, although the higher-dimensional gauge symmetry is globally broken by the 
compactification procedure, it always remains locally unbroken. Any local (sensitive to the UV 
physics) mass term for the scalars is then forbidden by the gauge symmetry and the Higgs 
mass only has a non-local and UV-finite origin.

The Gauge-Higgs unification idea has been applied to various frameworks. In the original 
scenario \cite{Manton} a compactification on $M_4 \times S^2$ is proposed with an additional 
symmetry ansatz on the gauge fields: only spherically symmetric configurations are allowed.
As by-product of this ansatz the gauge fields have a non-vanishing flux on $S^2$. This flux 
breaks the rank-two gauge symmetry group down to $SU(2)_{EW} \times U(1)_{Y}$. Furthermore a 
negative Higgs mass square term appears that could be responsable for the spontaneous breaking 
down to $U(1)_{EM}$. However, in \cite{Manton} the question of the stability of flux configurations 
was not analyzed, as only the configuration with the lowest angular momentum mode has been 
considered. The inclusion of higher momentum states could, in principle, modify the vacuum 
structure and restore (or not) the original symmetry\footnote{A similar problem was detailled 
described in \cite{Alfaro:2006is} where the Olesen-Nielsen instability \cite{Nielsen}, in the 
context of $M_4 \times T^2$ compactification, was analytically and numerically solved.}.

Few years later the Gauge-Higgs unification idea has been applied to the framework of gauge 
theories in non-simply connected space-time. When the space is non-simply connected, 
zero field strength configurations do not necessarily imply flat connection configurations. 
In these scenarios, in fact, non integrable (gauge-invariant) phases, associated to non-trivial 
Wilson loops, appear. These phases can be interpreted, from the 4D point of view, as vevs of 
the extra-dimensional gauge field (i.e. scalar) components. The minimum of the tree-level 
scalar potential does not depend on these vevs and, consequently, these phases are just free 
parameters that describe equivalent (classical) vacuum configurations of the theory. This 
degeneracy is lifted at the quantum level \cite{Luscher:1982ma,Hosotani}. The quantum stable 
vacuum of the theory is obtained minimizing the one-loop effective potential. Depending 
on the matter content included in the specific model, the minimum of the scalar potential 
preserves or not the original symmetry group. If the minimum corresponds to vanishing phases 
(vevs) then the original symmetry is preserved. Conversely, if at the minimum some of the phases 
(vevs) are non-trivial then the gauge symmetry group is dynamically broken \cite{Hetrick:1989jk,
Mclachlan:1989pf,Burgess:1990zu}. This mechanism, conventionally known as the 
{\em Hosotani mechanism}, can be used to reproduce the spontaneous electroweak symmetry 
breaking in the context of Gauge-Higgs unification. Moreover, as the Wilson loop is a gauge-invariant 
non-local operator (with any power of the scalar components of the gauge fields) through this 
mechanism one obtains an operator for the Higgs mass that is automatically free from any 
UV-divergence \cite{Hatanaka,finiteHiggs}.

This idea has been widely investigated in the context of $5D$ compactifications on $M_4 \times  
S^1 (S^1/Z_2)$, with either flat \cite{fiveflat} or warped extra-dimension \cite{fivewarped}. 
Some work has been done also in the context of $6D$ compactifications (with or without orbifolds) 
\cite{sixorbifold,Hosotani:2004ka}. 
In all these models, the need of having compactification in presence of singularities \cite{orbifold}
is mainly motivated by the necessity of obtaining $4D$ chiral fermions, starting from 
higher-dimensional theories \cite{Witten:1983ux}. 

Beside orbifold compactification, it is well known that $4D$ chiral theories can be obtained 
by compactifying in the presence of a background field, either a scalar field ({\it domain wall} 
scenarios) \cite{Rubakov}, either gauge - and eventually gravity - backgrounds with non trivial 
field strength ({\it flux compactification}) \cite{Randjbar}.

The idea of obtaining chiral fermions in the presence of abelian gauge and gravitational 
backgrounds was first proposed by Randjbar-Daemi, Salam and Strathdee \cite{Randjbar}, 
on a $6D$ space-time with the two extra dimensions compactified on a sphere. The presence 
of a (magnetic) flux in the background, living in the extra-dimensions, can produce $4D$ 
chiral theory, the mass splitting between the two $4D$ chiralities being proportional to the 
field-strength of the stable background. This seminal idea was right away adapted to heterotic 
string constructions \cite{Gross} and it is still nowadays deeply used in the framework of 
intersecting branes scenarios \cite{Aldazabal:2000sa}. 

From the field theory point of view $6D$ compactification on $M_4 \times \mathcal{T}^2$ in 
the presence of a background flux, living in the extra-dimensions, has been studied in 
\cite{Alfaro:2006is,Salvatori:2006pb}. The typical framework one can consider is that of an 
$U(N)$ gauge theory in six dimensions, with a non-vanishing $U(N)$ background field strength living 
in the extra-dimensions. As it is well known \cite{Randjbar}, the presence of an extra-dimensional 
stable magnetic flux, associated to the abelian subgroup $U(1) \in U(N)$, induces chirality 
in four dimensions. However, there is no stable background flux associated to the non-abelian 
field strength, since the $SU(N)$ gauge field is a flat connection on $\mathcal{T}^2$. 
Consequently any non-vanishing non-abelian background field strength, introduced ab initio, 
can be gauged away \cite{Ambjorn:1980sm}. The numerical prove of this statement is however 
technically quite difficult, requiring to solve explicitly the Olesen-Nielsen instability on the torus. 
This was done, for the first time, in \cite{Alfaro:2006is} where the complete $4D$ tree-level scalar 
potential was numerically minimized including simultaneously (a sufficient number of) Kaluza-Klein 
and Landau heavy modes. 

Besides producing $4D$ chirality, the presence of a non-vanishing $U(1)$ flux also affects the 
non-abelian part of the group, $SU(N) \in U(N)$, being connected to a topological quantity, 
conventionally known as the {\em non-abelian 't Hooft flux} \cite{tHooft}, and producing 
interesting $SU(N)$ symmetry breaking patterns. While the $SU(N)$ trivial 't Hooft flux case 
has been deeply analyzed in the literature, the field theory analysis and the phenomenological 
applications of the non-trivial (non-abelian) 't Hooft flux has been explored only recently. 
In \cite{Alfaro:2006is} an effective field theory approach was used to explicitly show the 
$SU(2)$ classical symmetry breaking pattern and the resulting gauge-scalar spectrum, for 
both the trivial and non-trivial 't Hooft non-abelian flux. In \cite{Salvatori:2006pb} such 
analysis was extended and generalized to the $SU(N)$ case. Recently, then, the symmetry 
breaking pattern of models with the simultaneous presence of orbifold and non-abelian 't Hooft 
flux has been analyzed by \cite{vonGersdorff:2007uz}. Models with N=1 supersymmetry 
have been also considered in \cite{Abe:2008fi}.

The main motivation of this paper is to study, at one-loop level, the symmetry breaking patterns 
analyzed at tree-level in \cite{Alfaro:2006is,Salvatori:2006pb}. To do this we calculate the one-loop 
effective scalar potential in the presence of 't Hooft flux. In the case of trivial 't Hoof flux, one therefore 
reduces to the well known results already present in the literature (see for example \cite{Hatanaka} for a 
$6D$ example). There was, however, no calculation available up to now of how the Hosotani mechanism 
does work in the presence of non-trivial 't Hooft flux. This generalization is provided here. 


The paper is organized as follows. 
In section \ref{sectiongen} we summarize the main aspects of a $6D$ theory in the presence of a 
generic $U(N)$ background living in the extra-dimensions. The symmetry breaking patterns obtained 
in the case of trivial and non-trivial t' Hooft flux are analyzed and the tree-level gauge and 
scalar spectrum are derived.
In section \ref{sectionfer} we recall the main notions about chiral fermions in the presence of a 
background (magnetic) flux. We discuss the relation between 't Hooft flux and magnetic flux and 
we explicitly  write the spectrum for fermions in the fundamental and adjoint representation.
In section \ref{sectionloop} we calculate the one-loop effective potential contribution of gauge, 
scalar and fermionic sectors, for both trivial and non-trivial 't Hooft flux and then in section 
\ref{sectionpheno} we discuss some phenomenological issues. Finally in section \ref{conclusions} 
we state our conclusions. In Appendix A we explicitly calculate the $U(N)$ wave-functions in the 
fundamental representation while in Appendix B we present the general formalism for calculating 
the one-loop effective scalar potential using the Heat Function method.

%% file: parte_gen.tex
\section{$U(N)$ gauge theory on $\mathcal{M}_4 \times \mathcal{T}^2$}
\label{sectiongen}

Consider a $U(N)$ gauge theory on a $6D$ space-time\footnote{Throughout the paper, with $x$ 
and $y$ we denote the ordinary and extra coordinates, respectively. Latin upper case indices 
$M,N$ run over all the six dimensional space, whereas Greek and Latin lower case indices 
$\mu, \nu$ and $a,b$ run over the four ordinary and the two extra-dimensions, respectively.}
where the two extra dimensions are compactified on an orthogonal torus $\mathcal{T}^2$. 
To completely define a field theory on a torus one has to specify the periodicity conditions: 
that is, to describe how the fields transform under the fundamental shifts $y \rightarrow y + 
\ell_a$, with $\ell_a$ being the vectors identifying the fundamental lattice shifts along the 
$a$-circle of length $l_a$. 
Let's denote with $T_a$ the embeddings of these shifts in the fundamental representation 
of $U(N)$. The general periodicity conditions\footnote{We consider here exclusively the case 
of internal automorphisms. For the most general case of external automorphisms one can refer 
to \cite{Hebecker,Quiros}.} 
for the gauge field $A_M$, that preserve $4D$ Poincar\'e invariance, read:
\be
\mathbf{A_{M}} (x,y + \ell_a) = T_a (y) \,\mathbf{A_M} (x,y)\,T_a^\dagger(y)  + 
                               \frac{i}{g} T_a(y) \partial_M  T_a^\dagger(y) \, .
\label{pcsgauge}  
\ee
This equation is derived from the fact that while individual gauge fields may not be 
single-valued on the torus, any physical scalar quantity, like the Lagrangian, must be. 
The periodicity conditions in Eq.~(\ref{pcsgauge}) are usually referred as Scherk-Schwarz 
boundary conditions \cite{SS}.

The transition functions $T_a(y)$, hereafter simply denominated {\em twists}, in order to preserve 
the $4D$ Poincar\'e invariance, can only depend on the extra-dimensional coordinates $y$.
Consistency with the geometry imposes the following $U(N)$ consistency condition on the twists 
\cite{tHooft,Lebedev:1988wd}:
\be 
T_1 (y+\ell_2)\,T_2(y)\,\,= \,\, e^{i \theta} \, T_2 (y+\ell_1)\,T_1(y) \,.
\label{U(N)_cons_cond}
\ee
This condition is obtained imposing that the value of the gauge field $A_M(y_1+l_1, y_2+l_2)$ 
has to be independent on the path which has been followed to reach the final point 
$(y_1+l_1, y_2+l_2)$ from the starting point $(y_1,y_2)$, modulo a constant element of the 
center of the group, which, for $U(N)$, is a phase. 
%
%
One can easily verify that the inclusion of fields that transform in a representation sensitive to 
the center of the group, like for example the fundamental representation, imposes, in 
Eq.~(\ref{U(N)_cons_cond}), the additional constraint $\theta=0$. As we are interested in models 
with the simultaneous presence of fields in the adjoint and in the fundamental representation, 
throughout the paper we will impose the following $U(N)$ consistency condition: 
\be 
T_1 (y+\ell_2)\,T_2(y) \,\,= \,\, T_2 (y+\ell_1)\,T_1(y) \,.
\label{U(N)fundamental}
\ee

The $U(N)$ twist matrices can be, locally, decomposed as the product of an element $e^{i v_a(y)} 
\in U(1)$ and an element $\mathcal{V}_a (y) \in SU(N)$ as follows:
\be
T_a (y) \,\,=\,\,e^{i v_a(y)} \,\,\mathcal{V}_a (y)\,.
\label{Tasplit}
\ee
Using this parameterization, the consistency condition of Eq.~(\ref{U(N)fundamental}) can be 
splitted in the $SU(N)$ and $U(1)$ part, respectively:
\bea
\label{SU(N)_cons_cond}
e^{2 \pi i \frac{m}{N}} \, \mathcal{V}_1 (y+ \ell_2)\,\mathcal{V}_2(y) &=& 
\mathcal{V}_2 (y+\ell_1)\,\mathcal{V}_1(y) \\
\Delta_2 v_1(y) - \Delta_1 v_2(y)  &=&  2 \pi \frac{m}{N} \, , 
\label{U(1)_cons_cond}
\eea
with $\Delta_a v_b(y) = v_b(y+\ell_a) - v_b(y)$. The exponential factor in Eq.~(\ref{SU(N)_cons_cond}) 
is nothing else that the center of $SU(N)$. The integer $m=0,1,..,N-1$ (modulo N) is a gauge 
invariant quantity called the {\em non-abelian 't Hooft flux}\cite{tHooft}. Furthermore, it coincides with 
the value of a quantized abelian magnetic flux living on the torus, Eq.~(\ref{U(1)_cons_cond}), or, 
in other words, with the first Chern class of $U(N)$ on ${\cal T}^2$. 

\subsection{Boundary conditions vs background flux}

Up to here we have discussed the general properties of a $6D$ $U(N)$ gauge theory with 
Scherk-Schwarz boundary conditions. We are interested now to particularize the discussion 
considering the specific set of $U(N)$ gauge field configurations characterized by a constant 
(background) field strength, living in the extra-dimensions and pointing in an arbitrary direction 
of the gauge space. The physical relevance of these configurations will be immediately clear in 
the following subsections.

Let's expand the $U(N)$ gauge field, $\mathbf{A_M}$, in terms of the stationary background, 
$B_M$, and the fluctuation field, $A_M$, around it as:  
\be
\mathbf{A_M} (x,y) \,= \, \, B_M(x,y) \,+\,A_M(x,y) \, = \, B_a (y) \, \delta_{aM} \,+\,A_M(x,y) \, .
\ee 
The specific form of the background field in the previous equation is chosen to guarantee 
$4D$ Poincar\'e invariance. In the presence of such a background, the general Scherk-Schwarz 
periodicity conditions for the fluctuation and background fields read:
\bea
A_M (x,y+\ell_a) \, &=& \, T_a (y) \, A_M(x,y) \, T_a^\dagger(y) \,, \label{backperiod1}\\
B_b (y+\ell_a) \, &=& \, T_a (y)\, \left[B_b(y)\, +\, \frac{i}{g}\, \partial_b \right]\, T_a^\dagger(y) \, .
\label{backperiod2}
\eea
Following the definition of Eq.~(\ref{Tasplit}), we can write the periodicity conditions for the $U(1)$ 
and $SU(N)$ part of the fluctuation and background fields\footnote{We use the following conventions 
for the $U(1)$ and $SU(N)$ generators, $\lambda_0$ and $\lambda_k$: $\lambda_0=\one / \sqrt{2 N}$ 
and Tr$[\lambda_k\lambda_{k'}]=\frac{1}{2}\delta_{k k'}$, with $k,k'=1,2,\dots,(N^2-1)$.} 
respectively as:
\bea 
A_M^{(0)} (x,y+\ell_a) &=& A_M^{(0)} (x,y) \,, \nn \\
B_b^{(0)} (y+\ell_a) &=& B_b^{(0)} (y)\, + \, \frac{\sqrt{2 N}}{g} \,\partial_b v_a(y) \, , 
\label{back0boundary}
\eea
\bea
A_M^{(k)} (x,y+\ell_a) \, \lambda_k &=& \mathcal{V}_a(y) \, A_M^{(k)}(x,y) \lambda_k \, 
          \mathcal{V}_a^\dagger(y) \,, \nn \\
B_b^{(k)} (y+\ell_a)  \, \lambda_k &=& \mathcal{V}_a (y)\, \left[ B_b^{(k)} (y)\, \lambda_k \, + \,
          \frac{i}{g}\,\partial_b \right]  \,\mathcal{V}_a^\dagger(y) \,.
\eea
Notice however that neither the twists or the background flux are gauge invariant quantities and so the 
split between Eq.~(\ref{backperiod1}) and Eq.~(\ref{backperiod2}) is purely conventional.

In general, not all the possible choices of background fields and boundary conditions are compatible. To 
illustrate this, let's discuss the simplest case of an $U(1)$ gauge theory (or restrict to the $U(1)$ sector 
of the $U(N)$ theory) and consider a constant background field strength: 
\bea
B^{(0)}_{ab} (y) = \partial_a B^{(0)}_b - \partial_b B^{(0)}_a = \frac{\mathcal{F}}{g \mathcal{A}}  \quad , \quad 
B^{(0)}_a (y) = - \frac{\mathcal{F}}{2 g \mathcal{A}} \epsilon_{ab} \,  y_b  \label{connection}
\eea 
with $\mathcal{F}$ a dimensionless costant (flux) and $\mathcal{A}$ the area of the torus. Compatibility 
between Eq.~(\ref{connection}) and the boundary conditions of Eq.~(\ref{back0boundary}) force $v_a(y)$ 
to be of the form:
\bea
v_a(y) &=& \frac{\mathcal{F}}{2 \mathcal{A}} \epsilon_{ab} \, \ell_a \, y_b 
\eea  
where $\mathcal{F}= 2 \pi m$ from Eq.~(\ref{U(1)_cons_cond}).
It was shown by \cite{Hosotani}, that in the case of a $SU(N)$ gauge theory on $\mathcal{M}_4 \times 
\mathcal{S}^1$, starting from a compatible choice of background field and boundary conditions on the circle, 
it is always possible to go to a gauge in which either the twist is trivial or the background field is vanishing, 
the latter defined as the {\em symmetric gauge}. Moreover it was shown that in this gauge the $5D$ twist 
coincides with the Wilson loop and can be parameterized in terms of non-integrable, gauge invariant phase: 
the Scherk-Schwarz phase \cite{SS}. This quantity, in the gauge in which the twist is trivial, appears 
instead as a background field component and can be interpreted, from the $4D$ point of view, as non-vanishing 
vev for the $4D$ scalar (gauge) field. 

Similarly to what happens in the $5D$ case, it was shown in \cite{Salvatori:2006pb} that also for a $SU(N)$ 
guage theory on $\mathcal{M}_4 \times \mathcal{T}^2$ it is always possible to choose a gauge, namely the 
{\em symmetric gauge} in which the $SU(N)$ background field strength on the torus vanishes and the 
$SU(N)$ twist matrices are constant. Let's define the $U(N)$ Wilson line and Wilson loop around the 
$a$-circle, respectively, as:
\bea
W_a (y_f,y_i) &=& \mathcal{P} \exp \left\{ig \int_{y_i}^{y_f} dz^b B_b (z)  
            \right \} \, T_a(y) \, ,  \label{Wline} \\
W_a (y,y+\ell_a) &=& \mathcal{P} \exp \left\{ig \int_{y}^{y+\ell_a} \hspace{-0.3cm} dz^b B_b (z)  
            \right \} \, T_a(y) \, \equiv \, W_a \, ,
\label{Wloop}
\eea
where $\mathcal{P}$ stands for the path-ordered product. It is immediate to see that the $U(1)$ 
part of the twist automatically cancels in Eq.~(\ref{Wloop}) with the exponential part of the abelian 
background field, due to the condition of Eq.~(\ref{back0boundary}). Consequently in the symmetric 
gauge the following relations hold:
\bea
\left( \mathcal{V}_a (y) \right)_{sym} \equiv V_a = W_a (y,y+\ell_a) \, , \quad  && \quad 
\left( B_{ab}^{(k)} (y) \right)_{sym} = 0 \nn \,. 
\eea
\label{Wloopsym}
\noindent 
Being the trace of Eq.~(\ref{Wloop}) a gauge invariant and $y$-independent quantity, one consequently 
ends up, in the $6D$ case, with two independent non-integrable Scherk-Schwarz phases.


However, contrary to the lower dimensional case, in the $6D$ case, the symmetry of the classical 
vacua depends of an additional gauge invariant quantity, the 't Hooft non-abelian flux. The relation 
of the non-abelian 't Hooft flux and the existence of a background (abelian) magnetic flux can be 
immediately understood calculating the trace of the $U(1)$ part of the abelian background field 
strength and using the abelian periodicity condition of Eq.~(\ref{back0boundary}):
\bea 
 & &  \frac{g}{N}\int_{\mathcal{T}^2}\, \hspace{-0.3cm} d^2y \mathrm{Tr} \left[B_{12}(y) \right] =  
     g \int_{\mathcal{T}^2}\, \hspace{-0.3cm} d^2y \frac{ \left( \partial_1 B_2^{(0)}(y) - 
      \partial_2 B_1^{(0)}(y) \right) }{ \sqrt{2 N}} \, = \, \\ 
 & & \hspace{1cm} = \,  \int \hspace{-0.1cm} dy_2 \, \partial_2 v_1 (y) - \int \hspace{-0.1cm} dy_1 \, \partial_1 v_2 (y) 
 =  \left[ \Delta_2 v_1 (y) - \Delta_1 v_2(y) \right]  = \frac{2 \pi m}{N}  \,.\nn
\eea  
That is, the 't Hooft consistency condition of Eq.~(\ref{U(1)_cons_cond}) implies the quantization 
of the abelian magnetic flux in terms of the non-abelian 't Hooft flux $m$. 

\subsection{Trivial 't Hooft flux: $m=0$}

The spectrum can be easily discussed in the {\em symmetric gauge}. For the $m=0$ case, 
Eq.~(\ref{SU(N)_cons_cond}) tell us that the two $V_a$ matrices commute and consequently 
can be parameterized as:
\be
V_a= e^{2 \pi i (\alpha_a \cdot H)}  \quad  \quad , \quad  \quad 
\alpha_a \cdot H \equiv \sum_{\rho=1}^{N-1} \alpha_a^\rho H_\rho 
\label{V_a_m=0}
\ee
with $H_\rho$ the $(N-1)$ generators of the Cartan subalgebra of $SU(N)$. The periodicity condition, 
and consequently the classical vacua, are characterized by $2 (N-1)$ real continuous parameters, 
$0\leq \alpha_a^\rho < 1$. These parameters are non-integrable phases, which arise only in a 
topologically non-trivial space and cannot be gauged-away. When all the $\alpha_a^\rho$ are vanishing 
the initial symmetry is unbroken. At classical level $\alpha_a^\rho$ are undetermined. Their values are 
dynamically determined at the quantum level \cite{Luscher:1982ma,Hosotani} where a rank-preserving 
symmetry breaking can occur. This dynamical and spontaneous symmetry breaking mechanism is 
conventionally known as the {\em Hosotani mechanism}.
%
In order to write down the explicit expression for the (tree-level) mass spectrum of the $4D$ 
gauge and scalar components of the $6D$ gauge field one can introduce the Cartan-Weyl 
basis for the $SU(N)$ generators. In addition to the $(N-1)$ generators of the Cartan subalgebra, 
$H_\rho$, one defines $N (N-1)$ non diagonal generators, $E_r$, such that the following 
commutation relations are satisfied:
\be
\left[ H_{\rho}, H_{\sigma} \right] = 0 \quad , \quad 
\left[ H_{\rho}, E_r \right] = q^\rho_r E_r \, .
\label{qcharge}
\ee 
In this basis, the $V_a$ act in a diagonal way, that is 
\bea 
V_a H_\rho V_a^{\dagger} = H_\rho \quad & , & \quad V_a E_r V_a^{\dagger} = e^{2 \pi i \, 
                        (\alpha_a \cdot q_r) } \, E_r \,,
\eea
and the four-dimensional mass spectrum reads simply:
\be
m^2_{(k)} = 4 \pi^2 \sum_{a=1}^2  \left(n_a + \alpha_a \cdot q_k  
            \,\right)^2 \frac{1}{l_a^2}   \quad \quad  ,\quad n_a \in \mathbb{Z} \,,
\label{SS}
\ee
with $k$ here labeling the $(N^2-1)$ $SU(N)$ gauge (scalar) components. For a gauge (scalar) 
field component $A_M^\rho$, associated to a generator belonging to the Cartan subalgebra, 
$H_\rho$, one has $q_\rho = (0,...,0)$ and the spectrum reduces to the ordinary Kaluza-Klein 
(KK) one. For a gauge (scalar) field component $A_M^r$ associated to the non-diagonal 
generators, $E_r$, one has, instead, $q_r \neq (0,...,0)$ and the mass spectrum is 
consequently shifted by a factor proportional to the non-integrable phases $\alpha_a^\rho$.
When all the $\alpha_a^\rho\neq 0$, then only the gauge field components associated 
to the generators of the Cartan subalgebra are massless. Therefore, the symmetry breaking 
induced by the commuting twists, $V_a$, does not lower the rank of $SU(N)$. This result is the 
one generally reported by literature (see for example \cite{Hosotani:2004ka} for a 6D analysis).

One can easily generalize these results to the $U(N)$ case adding an extra diagonal generator, 
$H_0 = \one_N/\sqrt{2N}$. Obviously $H_0$ commute with all the twists $V_a$ and consequently 
$A_M^0$ always remains unbroken. The maximal symmetry breaking pattern that can be achieved 
in the $m=0$ case, for an $U(N)$ gauge theory is given by: 
\be
U(N) \sim U(1) \times SU(N) \rightarrow U(1) \times U(1)^{N-1} = U(1)^N.
\ee
This symmetry breaking mechanism is exactly the same Hosotani mechanism one is used 
to in a $5D$ framework.

\subsection{Non-trivial 't Hooft flux: $m \neq 0$}

In the $m\neq0$ case, the twists $V_a$ don't commute between themselves and so necessarily 
they induce a rank-reducing symmetry breaking. The most general solution of the consistency 
relation Eq.~(\ref{SU(N)_cons_cond}) can be parameterized as follows \cite{Ambjorn:1980sm,
GonzalezArroyo:1982hz,Salvatori:2006pb}: 
\be
V_1 = \omega_1 \,\,P^{s_1}\,Q^{t_1} \quad , \quad  V_2 = \omega_2 \,\,P^{s_2}\,Q^{t_2} \,.
\label{gaiarda}
\ee
Here $s_a, t_a$ are integer parameters taking values between $0,...,(N-1)$ (modulo $N$) and 
satisfying the following constraint: 
\be
s_1\,t_2 \,\,- \,\,s_2\,t_1\,\,=\,\, \mK\,.
\label{st_constraint}
\ee
$P$ and $Q$ are $SU(N)$ constant matrices given by
\bea
P \equiv P_{\NK} \otimes \one_{\mathcal{K}} \quad & , & \quad 
Q \equiv Q_{\NK} \otimes \one_{\mathcal{K}} \,.
\label{PQ_paraculi}
\eea
In the previous equations we defined $\mathcal{K} \equiv \mathrm{g.c.d.} (m, N)$, $\mK \equiv m/\mathcal{K}$ 
and $\NK \equiv N/\mathcal{K}$. The matrices $P_{\NK}$ and $Q_{\NK}$ are the following $\NK \times \NK$ matrices: 
\be
\left\{ \begin{array}{lcl}
\left(P_{\NK}\right)_{jk} & = & e^{i \pi \frac{\NK - 1}{\NK}} \,\, \delta_{j,k-1} \\
\left(Q_{\NK}\right)_{jk} & = & e^{- 2 \pi i \frac{(k-1)}{\NK}} \,\,e^{i \pi \frac{\NK-1}{\NK}}\,\, \delta_{jk}
\end{array} \right. \hspace*{2em} j,k\,=\,1,2,..., \NK,,
\label{PQ_def}
\ee
satisfying the conditions 
\bea
P_{\NK}\,Q_{\NK} = e^{-2 \pi i \frac{1}{\NK}} Q_{\NK} P_{\NK} \quad & , & \quad 
\left(P_{\NK}\right)^{\NK} = \left(Q_{\NK}\right)^{\NK} = e^{\pi i (\NK-1)} \, .
\label{PQ_properties}
\eea
When ${\mathcal K}=1$, then $\NK =N$ and $P$, $Q$ reduce to the usual elementary twist matrices 
defined by 't Hooft \cite{tHooft}.

The matrices $\omega_a$ are constant elements of $SU(\mathcal{K}) \subset SU(N)$. They commute 
between themselves and with $P$ and $Q$. Therefore $\omega_a$ can be parameterized in terms 
of generators $H_j$ belonging to the Cartan subalgebra of $SU(\mathcal{K})$:
\be
\omega_a \,=\,e^{2 \pi i\, (\alpha_a \cdot H)} \quad \quad , \quad \quad 
\alpha_a \cdot H \equiv \sum_{\rho=1}^{\mathcal{K}-1}\,\alpha_a^\rho \,H_\rho
\label{omega_piccola}
\ee
Here $\alpha_a^\rho$ are $2 (\mathcal{K}-1)$ real continuous parameters, $0\le\alpha_a^\rho <1$.
As in the $m=0$ case, they are non-integrable phases and their values must be dynamically 
determined at the quantum level producing a dynamical and spontaneous symmetry breaking. 

The $m\neq 0$ four-dimensional mass spectrum is easily obtained using the following basis 
\cite{Salvatori:2006pb} for the $SU(N)$ generators
\bea
\tau_{(\rho,\sigma)} (\Delta, k_\Delta) &=& \left\{ 
\begin{array}{lll}
\mathrm{if}\, \left\{\begin{array}{l}
\rho=\sigma \\
\Delta=k_\Delta=0
\end{array} \right.
 & \Rightarrow &\left(\sum_{i=1}^{\rho} \lambda^{\mathcal{K}}_{(i,i)} - 
   \rho \lambda^{\mathcal{K}}_{(\rho+1,\rho+1)}\right) \otimes \one_{\NK} \\ 
 & \\
\mathrm{else} & \Rightarrow & \lambda^{\mathcal{K}}_{(\rho,\sigma)} \otimes \tau^{\NK} (\Delta, k_\Delta) 
\end{array}
\right.
\label{base_fighetta}
\eea
where $\Delta,\,k_\Delta$ are integers assuming values between $0,\dots,(\NK-1)$ while the indices 
$\rho,\sigma$ take values between $1,\dots,\mathcal{K}$, excluding the case $(\Delta=k_\Delta=0, 
\rho=\sigma)$ in which $\rho$ takes values between $1,\dots,(\mathcal{K}-1)$. 
The matrices $\lambda^\mathcal{K}_{(\rho,\sigma)}$ and $\tau^{\NK}$ are $\mathcal{K} \times 
\mathcal{K}$ and $\NK \times \NK$ matrices, respectively, defined as:
\bea
\left(\lambda^{\mathcal{K}}_{(\rho,\sigma)}\right)_{\rho' \sigma'} &=& \delta_{\rho \rho'} 
     \delta_{\sigma \sigma'} \nn \\ 
\tau^{\NK}(\Delta,k_{\Delta}) &=& \sum_{k=1}^{\NK} \, e^{2 \pi i  \frac{k}{\NK}\,k_\Delta}\,
     \lambda^{\NK}_{(k,k+\Delta)}  \,.
\label{gen_nodiag}
\eea
The definition of $\lambda^{\mathcal{\NK}}_{(n,n')}$ comes straightforwardly.

In this basis, the $SU(\mathcal{K})$ generators that commute with $P$ and $Q$ are simply given 
by $\tau_{(\rho,\sigma)} (0,0)$. In particular, the generators belonging to the Cartan subalgebra 
of $SU(\mathcal{K})$ are given by $H^{\rho}=\tau_{(\rho,\rho)}(0,0)$. The following commutation 
relations are satisfied:
\bea
\left[ \tau_{(\rho,\rho)}(0,0) , \tau_{(\sigma,\sigma)}(0,0) \right] &=& 0 \quad , \quad 
\left[ \tau_{(\tau,\tau)}(0,0), \tau_{(\rho,\sigma)}(\Delta,k_\Delta) \right] =
      q^{(\rho,\sigma)}_\tau \tau_{(\rho,\sigma)}(\Delta,k_\Delta) \,. \nn 
\eea
The action of the twists $V_a$ on this basis is given by
\bea
V_a \, \tau_{(\rho,\sigma)}(\Delta,k_\Delta) \,V_a^\dagger \,& = &\, 
  e^{\frac{2 \pi i}{\NK}(s_a \Delta + t_a k_\Delta)\,+ \, 2 \pi i \,
  ( \alpha_a \cdot q_{(\rho,\sigma)}) }\,\, \tau_{(\rho,\sigma)} (\Delta,k_\Delta)  \,,
\label{cho_K}
\eea
and the four-dimensional mass spectrum takes the following form:
\be
\label{m_spectrum_dani}
m^2_{(\rho,\sigma)}(\Delta,k_\Delta) = 4 \pi^2 \sum_{a=1}^2 \left(n_a + \frac{1}{\NK}\,(s_a\,\Delta\,\,+\,\,t_a\,k_\Delta) \,+\,
 \alpha_a \cdot q_{(\rho,\sigma)} \right)^2 \frac{1}{l_a^2} \, , \quad  n_a \in \mathbb{Z}
\ee
Therefore, beside the usual KK mass term, there are other two additional contributions. 
The first one, quantized in terms of $1/\NK$, is a consequence of the non-trivial commutation 
rule of Eq.~(\ref{PQ_properties}) between $P$ and $Q$ that induces the $SU(N) \rightarrow 
SU(\mathcal{K})$ symmetry breaking. Since $s_a,\,t_a$ cannot be simultaneously zero, 
the spectrum described by Eq.~(\ref{m_spectrum_dani}) always exhibits some (tree-level) 
degree of symmetry breaking. Given a set of $s_a, t_a$ and for all the $\alpha^\rho_a=0$ 
(that is $\omega_a=1$), only the gauge bosons components associated to $\tau_{(\rho,\sigma)}
(0,0)$, the generators of $SU(\mathcal{K})$, admit zero modes. This is an explicit breaking. 
%
%
The second contribution to the gauge mass is associated to the $\omega_a$ degrees of freedom and it 
depends on the continuous parameters $\alpha^\rho_a$. For $\mathcal{K}>1$ and all the non-integrable 
phases $\alpha_a^\rho \neq 0$, the only massless modes correspond to the gauge bosons associated 
to the Cartan subalgebra of $SU(\mathcal{K})$. The symmetry breaking pattern induced by the 
$\omega_a$ produce a Hosotani symmetry breaking that does not lower the rank of $SU(\mathcal{K})$. 

The maximal symmetry breaking pattern that can be achieved for an $U(N)$ gauge theory with matter 
fields in the fundamental is, in the $m\neq 0$ case, given by:
\be
U(N) \sim U(1) \times SU(N) \rightarrow U(1) \times U(1)^{\mathcal{K}-1} = U(1)^\mathcal{K}.
\ee
Obviously, when $\mathcal{K}=1$ the $SU(N)$ subgroup is completely broken, the only unbroken 
symmetry being the $U(1) \in U(N)$. This symmetry breaking pattern has no analogous in $5D$ 
frameworks\footnote{Except by introducing additional symnmetry breaking mechanism as for 
example orbifods.} and it's peculiar of higher-dimensional models where (topological) fluxes 
can be defined. 

Two final comments on the spectrum properties are in order. First of all, it could appears from 
Eq.~(\ref{m_spectrum_dani}) that gauge boson (or scalar) masses depend on the specific choice of 
the two integer parameters $s_a, t_a$. However, one can explicitly prove that for a given $\mK$,
any possible choice of $s_a, t_a$, satisfying the constraint of Eq.~(\ref{st_constraint}) gives the 
same boson (scalar) masses. We will see that this property will hold at the one-loop level too.
As a second comment notice that in both the cases of trivial and non-trivial 't Hooft flux, the classical 
effective $4D$ spectrum depends on the gauge indices but it does not depend on the Lorentz ones. 
This implies that at the classical level the $4D$ scalar fields $A_a$, arising from the extra-components 
of a $6D$ gauge fields, are expected to be degenerate with the $4D$ gauge fields $A_\mu$ with the 
same gauge quantum numbers. As we will see in section \ref{sectionloop} this degeneracy can be 
removed at the quantum level.

%% file: fermions.tex
%
\section{Fermions in the fundamental and adjoint of $U(N)$}
\label{sectionfer}
%

In the previous section we discussed the relation between boundary conditions and the symmetry 
breaking mechanism and we saw how to deduce the tree-level gauge mass spectrum. We consider 
now the fermionic sector, reviewing how to introduce fermions and define $4D$ chirality in the 
presence of a $U(N)$ background flux.

Fermions transforming in the fundamental or in the adjoint representation of $U(N)$ obey 
the following periodicity conditions:
\be
\Psi (x,y + \ell_a) = T_a(y)\,\Psi (x,y)  \quad , \quad \Psi (x,y + \ell_a) = T_a(y) \,
\Psi (x,y)\, T^\dagger_a(y) 
\nn \, ,
\label{pcsfermions}
\ee
where $T_a(y)$ must be, for gauge invariance, the same twists defined in Eq.~(\ref{pcsgauge}). 

A Dirac spinor in six dimensions has dimension eight.  We can thus construct a $6D$ Dirac 
fermion starting from two Dirac $4D$ spinors, that is
\bea
\Psi_{6D} = \left( \ba{c} \psi \\ \chi \ea \right)  \,,
\eea
and, for definiteness make use of the following representation of the Clifford algebra 
\bea
\Gamma^\mu & = & \gamma^\mu \otimes \one_2  \quad , \quad 
\Gamma^5 = \gamma^5 \otimes i \, \sigma_1 \quad , \quad 
\Gamma^6 = \gamma^5 \otimes i \, \sigma_2 \, ,
\eea
where $\gamma^\mu$  and $\gamma^5$ are the $4D$ gamma matrices, for example in the Weyl 
representation and $\sigma_i$ are the usual Pauli matrices. In six dimensions, chirality can be 
defined by means of the matrix
\be
\Gamma_7 = \prod_M \Gamma^M = \gamma^5 \otimes \sigma_3 \quad , \quad 
            \mathcal{P}_{L,R} = \left( \frac{1 \mp \Gamma_7}{2} \right) 
\ee
so that a $6D$ chiral fermion takes the form
\be \label{LandR}
\Psi_L = \mathcal{P}_L \Psi_{6D} = 
         \left( \ba{c}\psi_L \\ \chi_R \ea \right) \quad , \quad 
\Psi_R = \mathcal{P}_R \Psi_{6D} =  
         \left( \ba{c} \psi_R \\ \chi_L \ea \right) \,,
\ee
with $\psi_{L,R}$ the usual $4D$ left- and right-handed Weyl spinors. From the previous equation 
it is evident that a $6D$ chiral fermion is composed by two $4D$ Weyl fermions with opposite 
$4D$ chirality. 

The Lagrangian for a $6D$ massless left fermion, in the fundamental and in the adjoint of $U(N)$ 
can be written, respectively, as
\bea
\msc{L}_f &=& i \, \ol{\Psi}_L \Gamma^M D_M \Psi_L  \quad , \quad 
      D_M = \partial_M \, - \, i\,g\,\delta_{M,a}\,B_a(y) \,, 
\label{6DfermionL} \\
\msc{L}_f &=& i \, \ol{\Psi}_L \Gamma^M \msc{D}_M \Psi_L  \quad , \quad 
      \msc{D}_M = \partial_M \, - \, i\,g\,\delta_{M,a} \left[\,B_a(y) , \,\cdot\, \right] \,. 
\eea
$D_M$ ($\msc{D}_M$) are the $6D$ covariant derivative in the fundamental (adjoint) representation, 
with respect to the $U(N)$ background. From Eq.~(\ref{6DfermionL}) it is straightforward to obtain 
the following Klein-Gordon type equations for the zero mode of the $4D$ Dirac spinors in the fundamental 
of $U(N)$: 
\bea
\left( \partial^2 - D_z D_{\bar{z}}  + \left[D_z , D_{\bar{z}} \right] \right) \chi_R &=& 0 
\label{KGleft} \\
\left( \partial^2 - D_z D_{\bar{z}}  \hspace{2cm} \right) \psi_L &=& 0 \label{KGright}
\eea
with $D_z = \left(D_5 - i\, D_6\right)$, 
$D_{\bar{z}} = \left(D_5 + i\, D_6\right)$ and the commutator being: 
\be
\left[ D_z , D_{\bar{z}} \right] = 2g\, B_{56}  = 
     2g\, B^{(0)}_{56} \lambda_0 \,+\, 2g \sum_{k=1}^{N^2-1} B^{(k)}_{56} \lambda_k \,. 
\label{commutatorFU}
\ee
The extra-dimensional derivative terms in Eqs.~(\ref{KGleft}, \ref{KGright}) can be interpreted as 
mass terms in four dimensions. Moreover, the presence of a non vanishing commutator introduces a 
mass splitting between the $4D$ fermions of opposite chirality that thus can not have, 
simultaneously, a massless $0$-mode state \cite{Randjbar}.

The equivalent equations in the adjoint representation can be obtained simply replacing $D_M$ 
with $\msc{D}_M$, the mass splitting between fermions of different chirality being now given by:
\be
 \left[ \msc{D}_ z, \msc{D}_{\bar{z}} \right] = 2g\,\left[B_{56},\,\cdot\,\right] = 
       2g \sum_{k=1}^{N^2-1} B^{(k)}_{56} \left[ \lambda_k, \,\cdot\,\right] \,. 
\label{commutatorAD}
\ee
Therefore, as expected, the mass splitting for fermions in the adjoint is sensitive only to the 
non-abelian part of the flux. 

As argued in the previous section, all stable $SU(N)$ background configurations are trivial 
(i.e. the non-abelian part of the magnetic flux is vanishing) while the abelian part is 
quantized and proportional to the 't Hooft flux $m$. The mass splittings for fermions in the 
fundamental and in the adjoint representation of $U(N)$ are consequently given by:
\be
\left[ D_z , D_{\bar{z}} \right] = \frac{4 \pi}{\mathcal{A}} \frac{m}{N} \qquad , \qquad 
\left[ \msc{D}_z , \msc{D}_{\bar{z}} \right]  =  0 \, 
\label{comm_rules_app}
\ee
The previous equations reflect the well known result that, for a non-vanishing 't Hooft flux, 
only fermions in the fundamental can be chiral while theories with only fermions in the adjoint 
of $U(N)$ must necessarily be vector-like. 

In short, in our context the presence of chirality, is directly related to the presence of the 't Hooft 
flux through the abelian magnetic flux. In the rest of this section we'll consider separately the 
fermionic spectra for $m=0$ and $m\neq 0$, emphasizing the relatively less studied latter case. 

\subsection{Fermions in the presence of trivial 't Hooft flux: $m = 0$}
%
For $m=0$ both the $SU(N)$ and $U(1)$ part of the twists, defined in 
Eqs.~(\ref{SU(N)_cons_cond},\ref{U(1)_cons_cond}), separately commute. This  means that it is 
possible to find a gauge, i.e. the {\em symmetric} gauge, where $\mathcal{V}^{sym}_a = V_a$ is 
a constant matrix and $v^{sym}_a = 0$. In this gauge, obviously, both the $SU(N)$ and the $U(1)$ 
background field strength vanish and no background magnetic flux is present. If the $SU(N)$ twist is 
not trivial, i.e. $V_a \neq \one$, then some of the original symmetry group is broken, as seen in the 
previous section, and the corresponding fermionic zero modes are lifted. However the $4D$ theory 
is not chiral. 

In fact, let's start for definiteness with a $6D$ chiral fermion, $\Psi_L$ in the fundamental of 
$SU(N)$. The interaction terms between the $4D$ Weyl spinors $\psi_L$ and $\chi_R$ can be written as
\be
(m_6+im_5)\bar{\psi}_L\chi_R + (m_6-im_5)\bar{\chi}_R\psi_L \,.
\ee
Therefore, identifying $\psi_L$ and $\chi_R$ as the two chiral components of a $4D$ Dirac KK 
state, one obtains the following masses for the k$^{th}$ component of the fundamental multiplet:
\be
m_{n(k)}^2 = m_5^2+m_6^2 = 4 \pi^2 \sum_{a=1}^2 \frac{1}{l_a^2}(n_a + \alpha_a \cdot q_k)^2 
    , \quad n_a \in \mathbb{Z} \,, 
\ee
with $H^j \Psi_{(k)} = q^j_k \Psi_{(k)}$. 
In the case of vanishing 't Hooft flux, there is no difference in the mass spectrum between 
fermions belonging to the fundamental or the adjoint of $U(N)$, other than the difference in 
the charges $q_k$.

%


\subsection{Fermions in the presence of non-trivial 't Hooft flux: $m \neq 0$}
\label{secferm2}


Setting a non-trivial $SU(N)$ 't Hooft flux, along with a non-trivial $U(1)$ background, provides the 
conditions to have a chiral theory. Let's consider, then, fermions in the fundamental representation 
of $U(N)$. In this case the abelian magnetic flux allows us to distinguish left from right-handed 
fermions through a splitting of their extra-dimensional energy. As seen before this translates into a 
splitting of the $4D$ masses of the lowest modes with different $4D$ chirality. 

Four-dimensional masses for fermions in the $SU(N)$ fundamental representation are given by the 
eigenvalues of the extra-dimensional operators, with eigenfunctions consistent with the imposed periodicity 
conditions:
\bea
\label{KGleft1}
\left( - D_z D_{\bar{z}}  + \left[D_z , D_{\bar{z}} \right] \right) \chi_R^p &=
& m^2_{p(R)}\chi_R^p  \,\quad, \,\quad \chi_R^p(y + \ell_a) = T_a(y) \chi_R^p(y) \\
\left( -D_z D_{\bar{z}}  \hspace{2cm} \right) \psi_L^p &=& m^2_{p(L)}\psi_L^p \label{KGri
ght1} \quad\, ,\, \quad \psi_L^p(y + \ell_a) = T_a(y) \psi_L^p(y) 
\label{fundfereq}
\eea
One should notice that while the operators act diagonally in the N-dimensional gauge space, the 
$N\times N$ matrices appearing in the boundary conditions are not diagonal and consequently 
they mix different components within the multiplet. With the following definition of creation and 
annihilation operators \cite{Giusti:2001ta}:
\be
a^{\dagger} =- \sqrt{\frac{N\mcl{A}}{4\pi m}}D_z  \quad , \quad a = \sqrt{\frac{N\mcl{A}}{4\pi m}} D_{\bar{z}}
\ee
it is immediate to show that the energy eigenstates are equally spaced, differing only in the presence 
of the zero mode for the case of the left-handed field. Diagonalizing the $4D$ Lagrangian the following 
mass spectrum is obtained
\be
m^2_{p(R)} = \frac{4\pi m}{\mcl{A}N}(p+1) \quad , \quad m^2_{p(L)} = \frac{4\pi m}{\mcl{A}N} p 
\qquad {\rm with} \qquad p \in \mathbb{N} 
\label{ferspec}
\ee
that is, there is no massless eigenstate for the right-handed fermion. Notice the important fact that 
the Scherk-Schwarz phases are completely absorbed and don't show up in the spectrum. This ultimately 
means that fermions in the fundamental in the presence of flux will not help in solving the vacuum 
degeneracy.

One apparent oddity is the fact that now there seems to be $N$ solutions to the equations, 
one for each direction of the $SU(N)$ fundamental. However, we know that the remaining 
symmetry after the breaking is, at most, $SU(\mcl{K})$. For the case $\alpha_1 =\alpha_2 = 0$ 
it is not obvious how those $N$ fermions arrange themselves in $SU(\mcl{K})$ representations. 
Ultimately it is proven solving directly the equations, that only $\mcl{K}$ independent 
degrees of freedom remain from the original $N$. These indeed organize in the fundamental of 
$SU(\mcl{K})$. The full solution can be found in Appendix \ref{AppendixA}.

Finally we can address the possibility of having adjoint fermions. Clearly, these fermions 
are as ``blind" to the 't Hooft flux as the gauge fields. For them, the matrices $V_i$, now 
written in the adjoint representation of $SU(N)$ commute, again due to the fact that this 
representation is not faithful. They will be generated by some element of the Cartan 
subalgebra $\alpha_a\cdot H$ of $SU(N)/Z_N$ which will also give rise to a Scherk-Schwarz 
like mass term for KK tower
\be
\left. m_{adj}^2\right._{(\rho,\sigma)} = 4\pi^2 \sum_{a=1}^2\left( n_a + \frac{1}{\NK}(s_a\Delta +t_ak_\Delta) + 
       \alpha_a \cdot q_{(\rho,\sigma)} \right)^2 \frac{1}{l^2_a}
\label{m_fermion_flux}
\ee
The symmetry group that remains is rank $(\mcl{K}-1)$ and depends on the values of $\alpha_a$. 
Notice that the fermions will arrange themselves in representations of the resulting group. In particular, 
if we start from fermions in the adjoint of $SU(N)$ and $\alpha^a = 0$, we end up with $\NK^2$ 
adjoint representations and $(\NK^2-1)$ trivial representations of $SU(\mcl{K})$ in the compactified 
theory. Also for the fermionic spectrum one can explicitly verify that for a given $\mK$,
any possible choice of $\{s_a, t_a\}$, satisfying the constraint of Eq.~(\ref{st_constraint}) gives the 
same fermion masses. We will see that these properties will hold at the one-loop level too.



%% file: one_loop_new.tex
%
\section{One-loop effective potential on $\mathcal{M}_4 \times \mathcal{T}^2$}
\label{sectionloop}
%

The favored approach for the calculation of the one-loop effective potential in the extra-dimensional framework \cite{Hosotani, fiveflat} has been the direct computation through the master formula
\be \label{master-formula}
V_{eff} = \frac{i}{2}\trm{Tr}\ln\trm{Det} \left(D_M D^M \right) \,.
\ee
The effective potential is obtained as a sum over all the eigenvalues of the quadratic $(4+d)$-dimensional 
operator, $D_MD^M$. Usually this entails an integral over continuous four-dimensional eigenvalues 
as well as a discrete sum over extra-dimensional ones.

In this work, instead, we will compute the one-loop effective action for a $U(N)$ gauge theory on $\mathcal{M}_4 \times \mathcal{T}^2$ using the heat kernel technique\footnote{For an 
approach similar to the one used in this paper see for instance \cite{Hetrick:1989jk}.}. A brief introduction containing the main formulas is given in the appendices. The generality of this method permits computing directly in the complete higher-dimensional manifold rather than performing the dimensional reduction and summing over the resulting $4D$ degrees of freedom as is usual. In some circumstances, in particular when discussing the ultraviolet properties of the theory, this is crucial \cite{Alvarez:2006we} and to some extent has motivated our choice.

Since the heat kernel computation takes place explicitly in coordinate space, it results in a very useful instrument to distinguish contributions from local (ultraviolet sensitive) and non-local (ultraviolet insensitive) diagrams. The local contributions do not depend on the periodicity conditions and are invariant under all the original symmetries. Thus, they do not contribute to determine the symmetry breaking order parameters. Only non-local contributions will be relevant for symmetry breaking, which is then protected from ultraviolet divergences. 

In any case we have found that, at least in the case of vanishing 't Hooft flux, the non-local pieces of the effective potential, computed in the complete manifold and in the reduced theory, do coincide. We find no reason to expect a change in this picture when adding non-trivial 't Hooft flux. 

The details of the whole procedure are given in the appendices. For the main purposes of the following sections, it is enough to quote here the final result. After regularization, one obtains the following contributions to the one-loop effective action:
\begin{itemize}
\item Gauge bosons and ghosts:
\begin{eqnarray}
\Gamma_{(1)}^{g+gh}  &=& - 4\,\frac{V^{4+2}}{\pi^{3}} \,\sum_{w_1,w_2\neq0}\,\frac{\mathrm{Tr} 
 \,\left(\,W_1^{w_1}\,W_2^{w_2} \,\right)}{\left[ (l_1 w_1)^2 + (l_2 w_2)^2\right]^{3}} \,.
 \label{vec_ghost_extra}
\end{eqnarray}
The overall factor $4$ is due to the fact that for a flat manifold and gauge background with zero field-strength, the only effect of the ghosts is to reduce to four the possible polarizations of a $6D$ gauge boson\footnote{The general quadratic fluctuation operators for gauge bosons and ghosts are 
\bea
\mathrm{gauge} &\rightarrow& g_{\mu \nu} D^2 + R_{\mu \nu} -2 i g F_{\mu \nu} \,\,, \nonumber \\
\mathrm{ghosts} &\rightarrow& D^2 \,\,. \nn
\eea}.
\item Matter fields in the representation $\mathcal{R}$ of $U(N)$ 
\begin{eqnarray}
 \hspace*{-0.5cm}\Gamma^{f,s}_{(1)}   & = & - \eta_{f,s} \,
 \frac{V^{4+2}}{\pi^{3}} \,\sum_{w_1,w_2 \neq 0}\,\frac{\mathrm{Tr}_{\mathcal{R}} 
 \,\left(W_1^{w_1}\,W_2^{w_2}\,\right)}{\left[ (l_1 w_1)^2 +(l_2 w_2)^2 \right]^{3}}  \,,
\label{effpotmatter}
\end{eqnarray}
where $\eta_f = -4$ and $\eta_s = 2$ for Weyl fermions and complex scalars respectively. 
\end{itemize}
Here, $V^{4+2}$ is the $6D$ volume, $\mathrm{Tr}$ denotes the trace over the chosen $U(N)$ representation and $W_a \equiv W_a(y,y)$ is the Wilson loop. We also find that fields in representations sensitive to the 't Hooft flux, as for example the fundamental one, don't help in removing the degeneracy among the infinity of $U(N)$ vacua. This can be clearly seen for fermions in the fundamental representation if one observes that the spectrum (\ref{ferspec}) does not contain any dependence on the SS phases appearing in the background and in the periodicity conditions. This in turn implies that the one-loop effective action is independent of such parameters and therefore the contribution is only a (divergent) constant, that is, vacuum energy.  

Summarizing, only representations for which the commutator of covariant derivatives is zero help in removing the degeneracy among the infinity of $U(N)$ vacua. While in the case of trivial 't Hooft flux, $m=0$, all representations fall in this category, for non-trivial 't Hooft flux, $m\neq0$, only representations insensitive to the center of the $U(N)$ gauge group influence the determination of the true vacuum. 
  
\subsection{The $m\neq0$ case in detail}

We concentrate now on the one-loop effective potential for the case of non-trivial 't Hooft flux, $m\neq0$. The main purpose here is to use the general formulas previously derived and point out similarities and differences with respect to the case, commonly treated in the literature, of trivial 't Hooft flux $m=0$. In order to simplify the discussion, the background {\em symmetric} gauge is used. In such a gauge, indeed, the vacuum gauge configurations are trivial and the $SU(N)$ part of the twists are constant matrices coinciding with the Wilson loops, see Eq.~(\ref{Wloopsym}). 

It is possible to show that the discrete part of the Wilson loops only affects the overall scale of the one-loop effective action but not its shape. Consider for example the contribution due to gauge and ghost fluctuating fields. In this case, the trace appearing in Eq.~(\ref{vec_ghost_extra}) can be reduced to
\be
\trm{Tr} \left[ \,V_1^{w_1} \,V_2^{w_2} \,\right] =  \sum_{\rho,\sigma} \omega_1^{w_1}\omega_2^{w_2} \cdot
\sum_{k_\Delta, \Delta} e^{\frac{2\pi i}{\tilde{N}}[(s_1w_1+s_2w_2)\Delta + (t_1w_1+t_2w_2)k_\Delta] } \,.
\ee
Furthermore one can easily prove that: 
\bea \displaystyle
\sum_{k_\Delta, \Delta} e^{\frac{2\pi i}{\tilde{N}}[(s_1w_1+s_2w_2)\Delta + (t_1w_1+t_2w_2)k_\Delta] } 
\, &=& \,  
\left\{ \ba{lc} 
\NK^2 \hspace{0.5cm} & \trm{if~} w_1 = \NK n_1, \, w_2 = \NK n_2 \\
0                    & \trm{otherwise} 
\ea \right. \nn
\eea
Therefore the effective potential contribution for gauge and ghosts is simply:
\be
\Gamma_{(1)}^{g+gh}  = - 4\NK^2 \frac{V^{4+2}}{\pi^{3}} \,\sum_{n_1,n_2\neq0}
\frac{\mathrm{Tr} \,\left[\omega_1^{\NK n_1 } \,\omega_2^{\NK n_2} \right]}
{\left[ (\NK l_1 n_1)^2 + (\NK l_2 n_2)^2\right]^{3}} \,.
\label{vec_ghost_extra_SU(K)}
\ee
From the previous result one can notice that the effective potential depends only on the continuous parameters contained in the twists and on $\mK$, but it does not depend on the specific choice made for the discrete parameters $s_a, t_a$ compatible with the constraint Eq.~(\ref{st_constraint}). Consequently, the resulting one-loop gauge mass spectrum will depend only on the value of the SS phases and on $\mK$. Also at one-loop level, two different sets of $s_a, t_a$ (for a fixed $\mK$) represent only different parameterizations of the same vacuum. Concerning gauge and ghost contributions, Eq.~(\ref{vec_ghost_extra_SU(K)}) shows clearly also that, apart from an overall scale, a $U(N)$ theory with non-trivial 't Hooft flux $m$ coincides with the case of a $U(\mathcal{K}) \subset U(N)$ theory on a torus with lengths given by $L_a = \NK l_a$ and with commuting periodicity conditions given by $\omega_a^{\NK}$. This is the expected symmetry according to the previous tree-level analysis. 

\subsection{Reducible adjoint representations}

Finally, in this subsection, we want to exemplify the results previously obtained focusing on those 
representations that are not sensitive to the center of the group, considering in particular tensor 
products of adjoint representations. 
Let $i_R,j_R,\dots$ be indices running from $1$ to the dimension $R$ of such representation. 
With $|i_R\rangle$ we represent the states that diagonalize the action of the boundary conditions $V_a$. 
In the case of the adjoint representation, the indices and the parameters appearing in Eq.~(\ref{cho_K}), 
$\rho, \sigma, \Delta, k_\Delta$, are now functions of $i_R$ and the equation has consequently 
to be rewritten as: 
\be
V_a |i_{adj} \rangle = \exp \left\{ \frac{2\pi i}{\NK} ( s_a\Delta^{(i)} + t_a k_\Delta^{(i)} ) + 
2\pi i \left( \alpha_a \cdot q_{(\rho_{i}, \sigma_{i} )}\right)   \right\} |i_{adj} \rangle \,.
\ee
The important fact is that the action of the $V_a$ over the product of any number of adjoint 
representations is already diagonal because of this choice of basis. If we take, for definiteness, 
the case of the product of two adjoint representations it turns out that, although clearly we can not 
say which combination of $|i_{adj} \rangle |j_{adj} \rangle$ belongs to this or that irreducible 
representation, we can nevertheless say how they transform under the simple diagonal action 
of the $V_a$, namely:
\bea
V_a(|i_{adj}\rangle|j_{adj}\rangle) &=& V_a|i_{adj} \rangle \times V|j_{adj}\rangle \nonumber \\
& = & \exp \left\{ \frac{2\pi i}{\NK} \left( s_a (\Delta^{(i)} + \Delta^{(j)}) + 
      t_a (k_\Delta^{(i)} + k_\Delta^{(j)} )\right)  \right. \nonumber \\
&&\hspace{4ex}+ \Bigg. 2\pi i \left( \alpha_a \cdot ( q_{(\rho_{i}, \sigma_{i} )} + 
      q_{(\rho_{j}, \sigma_{j} )})\right)   \Bigg\} |i_{adj}\rangle |j_{adj}\rangle \,.
\eea
In other words, we can obtain the spectrum without the need of identifying each field with its 
irreducible representation. The spectrum for the matter fields belonging to the product of 
two adjoint representations reads:
\bea \label{new-masses}
m_{i,j}^2 &=& 4\pi^2\sum_{a=1}^2 \Big( \Big. n_a +  \frac{1}{\NK}(s_a(\Delta^{(i)} + 
\Delta^{(i)} ) + t_a(k_\Delta^{(i)} + k_\Delta^{(j)}) ) + \nonumber  \\
&& \hspace{8ex}+ \alpha_a \cdot(q_{(\rho_{i},\sigma_{i})} + q_{(\rho_{j},\sigma_{j}) } ) \Big. \Big)^2 \, .
\eea
After the symmetry breaking, the residual symmetry of the theory is $SU(\mcl{K})$. The masses 
coming from each $SU(\mcl{K})$ representation can be clearly identified by their weights, formed 
by adding the weights of the adjoint $q_{(\rho_{i},\sigma_{i})} + q_{(\rho_{j},\sigma_{j}) } 
= q'_{ij}$. For each $q'_{ij}$ there are $\NK^2$ fields.

The effective potential is a function of the mass eigenvalues alone. Eq.~(\ref{new-masses}) tells us 
that in order to find the effective potential for a product of adjoints it is sufficient to substitute in 
the final contribution of a particular field
\bea
\Delta & \rightarrow & \Delta^{(i)} + \Delta^{(i)} \\
k_\Delta & \rightarrow & k_\Delta^{(i)} + k_\Delta^{(j)} \\
q_{(\rho,\sigma)} & \rightarrow & q'_{ij}
\eea
and sum over all contributions. This can be implemented rigorously in our formalism by allowing the 
Green function to wear two pairs of gauge indices, one pair for each adjoint representation. The main 
point here is that, although we have in fact deduced the contribution to the effective potential of 
the reducible representation formed by the product of two adjoints, one can identify each one of the 
terms with one of the irreducible components through their weights. Therefore we argue that the contribution 
to the effective potential of any irreducible representation that can be obtained as a component of some 
product of adjoints is completely determined by the representation weights and given by the formula 
Eq.~(\ref{vec_ghost_extra_SU(K)}) where the $\omega$'s carry the weight information.

\subsection{Scalar fields mass splitting}
\label{scalarsplit}
An interesting aspect of the one-loop analysis is related to the radiative contribution to the masses of the 
$4D$ scalars that arise from the extra components of the gauge fields. It is well known \cite{Alfaro:2006is} 
that gauge ($A_{\mu}$) and scalar ($A_{a}$) masses obtained through a non singular toroidal compactification 
are degenerate. In particular, regardless of the Lorentz indices, the square masses are given by the 
eigenvalues of the operator 
\bea
m^2 \equiv - \msc{D}^2 = - \left(\msc{D}_1^2 + \msc{D}_2^2 \right) \,,
\label{mass_pheno}
\eea
where $\msc{D}_a$ are the covariant derivatives with respect to a fixed stable background compatible 
with the periodicity conditions. As seen before, the covariant derivatives for an adjoint representation 
always satisfy $\left[\msc{D}_1,\msc{D}_2\right]=0$.

The fact that the operator in Eq.(\ref{mass_pheno}) does not depend on the $4D$ Lorentz indices, 
implies that in the $4D$ effective theory, one should always find at least a scalar degenerate 
with any gauge field. The discussion of the scalar masses, however, is a delicate issue and it needs 
some additional comments. 

In case of an unbroken gauge symmetry the extra-dimensional (gauge or scalar) fields $A_a$ can be expanded 
in terms of usual Kaluza-Klein modes:
\bea
A_a (x,y)\,=\, \frac{1}{\sqrt{\mathcal{A}}}\,\sum_{n_1,n_2=-\infty}^{\infty}\,A_{\vec{n}, \, a} (x) \, e^{2 \pi i \left( \frac{n_1}{l_1} y_1 + \frac{n_2}{l_2} y_2\right)} \,,
\eea
Integrating over the torus surface, one obtains a mass term for the a specific combination of the $4D$ 
scalar degrees of freedom:
\bea
m^2_{(\vec{n},k)} \,A^{(-\vec{n},k)}(x)\, A^{(\vec{n},k)}(x) \, = \, 
\left(m^2_{(n_1,k)}+m^2_{(n_2,k)}\right) \,A^{(-\vec{n},k)}(x)\, A^{(\vec{n},k)}(x) \, ,
\label{myspectrum}
\eea
with $k$ the index of the adjoint representation and $m_{(n_a,k)}=2 \pi n_a/l_a$ the usual KK mass term.
While the field $A^{(\vec{n},k)}(x)$ in Eq.~(\ref{myspectrum}), defined as:
\bea
A^{(\vec{n},k)} (x) \,=\,\frac{1}{\sqrt{m^2_{(\vec{n},k)}}} \,\left( 
 m_{(n_1,k)} A_2^{(\vec{n},k)} (x) - m_{(n_2,k)} A_1^{(\vec{n},k)}(x) \right) \, ,
\label{Adiagonal}
\eea
takes a KK mass, the orthogonal combination 
\bea
a^{(\vec{n},k)}(x) \,=\,\frac{1}{\sqrt{m^2_{(\vec{n},k)}}} \,\left( 
 m_{(n_1,k)} A_1^{(\vec{n},k)} (x) + m_{(n_2,k)} A_2^{(\vec{n},k)} \right) \,
\label{adiagonal}
\eea
remains massless.
The $4D$ scalars $a^{(\vec{n},k)}(x)$ are coupled to the $4D$ gauge fields by a derivative 
coupling proportional to:
\bea
g \int_{\mathcal{T}^2}\hspace{-0.25cm}  d^2y\,\,A_\mu \partial^\mu \, 
\left( \msc{D}_1 A_1 + \msc{D}_2 A_2\right) \,= \,g \,\sum_k\,\sum_{n_1,n_2=-\infty}^{\infty} 
\hspace{-0.25cm} \sqrt{m^2_{(\vec{n},k)}}\,A_{\mu}^{(-\vec{n},k)} \, \partial^\mu \,a^{(\vec{n},k)} \,.
\eea 
Having the quantum numbers of the current associated to the broken gauge symmetry the scalars 
$a^{(\vec{n},k)}$ can be seen as the pseudo-Goldstone bosons associated to the compactification 
symmetry breaking (from $6D$ to $4D$). The fields $a^{(\vec{n},k)}$ with $n\neq 0$ are absorbed 
by the corresponding KK gauge bosons that acquire a KK mass term leaving unchanged the counting 
of total degrees of freedom. 

In case of non-trivial boundary conditions these formula can be straightforwardly modified and the 
corresponding mass terms, $m_{(n_a,k)}$, read from Eq.~(\ref{SS}) or Eq.~(\ref{m_spectrum_dani}) 
depending on the value of the 't Hooft flux $m$. Notice that now the index $k$ in  Eqs.~(\ref{myspectrum})-(\ref{adiagonal}) runs over the indices of the Cartan-Weyl basis of 
Eq.~(\ref{qcharge}) for the $m=0$ case, while for the $m\neq 0$ case, $k$ represents the set of 
indices $(\Delta, k_\Delta, \rho, \sigma)$ characterizing the basis in Eq~(\ref{cho_K}). For any broken 
symmetry there is a physical scalar field with a mass $m^2_{(\vec{n},k)}=m^2_{(n_1,k)}+m^2_{(n_2,k)}
\neq 0$, degenerate with the associated gauge boson plus a massless pseudo-Goldstone boson. Instead, 
for gauge and scalar fields associated to generators of conserved symmetry, $m^2_{(0,k)}=0$, and 
consequently there are two massless (and physical) scalars, $A^{0,k)}(x)$ and $a^{0,k)}(x)$ degenerate 
with the associated gauge field. In the $m=0$ case, these zero modes arise from the scalars associated 
to the generators of the $SU(N)$ Cartan sub-algebra while in the $m\neq0$ case they are associated to 
the generators of the Cartan sub-algebra of $SU(\mathcal{K}) \in SU(N)$. 

However, the presence of such massless scalar degrees of freedom is, in general, an unwanted feature 
for obvious phenomenological reasons. Luckily, these scalars associated to the conserved symmetries 
receive a mass term from loop contributions. One can directly check this fact by taking the second 
derivative of the effective potential with respect to the continuous SS parameters $\alpha_i$ and 
evaluating it at the minimum. 
The reason why these masses are not forbidden by gauge invariance can be seen by writing all the 
gauge invariant effective operators that can appear at one-loop level. Let's work for definiteness in the 
{\em symmetric gauge}. Then the fields $A^{(k)}(x,y)$, with $k$ belonging to the Cartan sub-algebra of 
$SU(N)$ (or $SU(\KK)$ if $m\neq0$) are periodic in the extra dimensions. Gauge transformations 
 $U = e^{i \beta \cdot H}$ with $\beta_k(x,y)$ a periodic function in the y-coordinates preserves 
the residual guage invariance
\be
A_a^{(k)}(x,y) \rightarrow (U A_a(x,y) U^\dagger)_k + \frac{i}{g}(U\partial_a U^\dagger)_{k} = 
A_a^{(k)}(x,y) - \frac{1}{g}\partial_a \beta_k(x,y)
\label{ginv}
\ee
Now, the following class of operators
\be 
O_n = c_n\trm{Tr}\left(\int dy_1dy_2 A_a^{(k)}(x,y)\right)^n \qquad \qquad \forall n\in\mbb{N}
\label{inv-op}
\ee
are gauge invariant for any tranformation Eq.~(\ref{ginv}) with periodic $\beta_{k}(x,y)$. In particular, 
the operator with $n=2$, represents a gauge invariant mass term for the scalar fields. 
So, while in the tree-level Lagrangian, locality and gauge invariance forbid any mass terms for 
the $6D$ gauge bosons at one-loop order, instead, new non-local and gauge-invariant operators 
appear in the effective action, some of them playing the role of $4D$ scalar mass terms. 
For this to happen it is fundamental to work with non-simply connected manifolds. In the case of 
a space-time of the type $\mathcal{M}_4 \times \mathcal{T}^2$, the non-local operators are 
associated to the non-contractible cycles of $\mathcal{T}^2$ and they can only contain the 
extra-components of a $6D$ gauge boson,  $A_a$. Therefore, only these can take a mass 
whereas the ordinary components $A_\mu$ do not. 

%% file: pheno_new.tex
%
\section{Some Phenomenology}
\label{sectionpheno}
%

In order to make more explicit the previous statements, we are going to discuss here the particular example 
of a symmetry breaking patter $SU(N) \rightarrow SU({\mathcal{K}})$ with $\mathcal{K}=2$. Adopting the 
standard notation used in the literature \cite{sixorbifold}, where only the $m=0$ case was treated, one 
can rewrite the one-loop effective potential contribution to gauge and ghost $\Gamma^{g+gh}$ of 
Eq.~({\ref{vec_ghost_extra_SU(K)}) as:
\bea 
\Gamma_{(1)}^{g+gh}  &=& - 8\NK^2 \,\frac{V^{4+2}}{\pi^{3}} \left\{2 \hspace{-0.25cm} 
                 \sum_{n_1,n_2=1} \frac{\cos (2\pi \NK n_1\alpha_1 )\,\cos (2 \pi \NK n_2\alpha_2 )}
                {\left[ (\NK l_1 n_1)^2 + (\NK l_2 n_2)^2\right]^{3}} \, \, + \, \right. \nonumber \\
& & + \, \left.\sum_{n_1=1} \frac{\cos (2\pi \NK n_1 \alpha_1 )}{ (\NK l_1 n_1)^6} +\sum_{n_2=1} \frac{\cos (2\pi \NK n_2\alpha_2 )}{ (\NK l_2 n_2)^6} \right\} \,=\, 2 \, \Gamma^{q=1}_{(1)} ,
\label{vec_ghost_SU(K)_literature}
\eea
with the weights for the adjoint of $SU(2)$ equal to $\pm 1$. As one expect, for an $SU(2)$ gauge group, 
the effective potential depends from the two SS (continuous) parameters: $\alpha_1, \alpha_2$. The only 
remnant of the original group and of the symmetry breaking driven by the non-trivial 't~Hooft flux $m$ 
is the presence of the coefficient $\NK=N/{\mathcal{K}}$. This term modifies the periodicity of the 
effective action and consequently it may change the location of the stable one-loop minima of the effective 
potential. In fact, the effective potential in Eq.~(\ref{vec_ghost_SU(K)_literature}) is invariant under 
the following transformations:
\bea
\alpha_a & \rightarrow & \alpha_a + k_a/\NK \, ,
\eea
From the inspection of Eq.~(\ref{vec_ghost_SU(K)_literature}) one obtains that the stationary conditions 
for $\Gamma_{(1)}^{g+gh}$ are given by $2  \,\alpha_a  \NK  = k_a$ ($k_a \in  \mathbb{Z}$) with the 
minima identified by $(\alpha_1,\alpha_2) = (k_1/\NK,k_2/\NK)$. 
In Fig.~(\ref{fig1}) we plot the gauge contribution to the effective action, $\Gamma_{(1)}^{g+gh}$, 
as function of $(\alpha_1,\alpha_2)$ for $\mathcal{K}=2$ and two different choices of $\NK$. The 
corresponding results for a $6D$ model with trivial 't Hooft flux can be easily obtained by setting 
$\NK=1$ (i.e. $L_a=\ell_a$) in Eq.~(\ref{vec_ghost_SU(K)_literature}). For example one can see that 
$\Gamma^{q=1}_{(1)}$ defined in Eq.~(\ref{vec_ghost_SU(K)_literature}) coincides with the function 
$I(\alpha,\beta)$ defined in Eq.~(4.17) of \cite{Hosotani:2004ka}, while the total contribution from gauge 
and ghost is obviously different as in  \cite{Hosotani:2004ka} the authors are considering a $\mathcal{T}^2/
\mathcal{Z}_2$ orbifold model.

\begin{figure}[!t]
\begin{center}
\begin{tabular}{cc}
  \epsfig{file=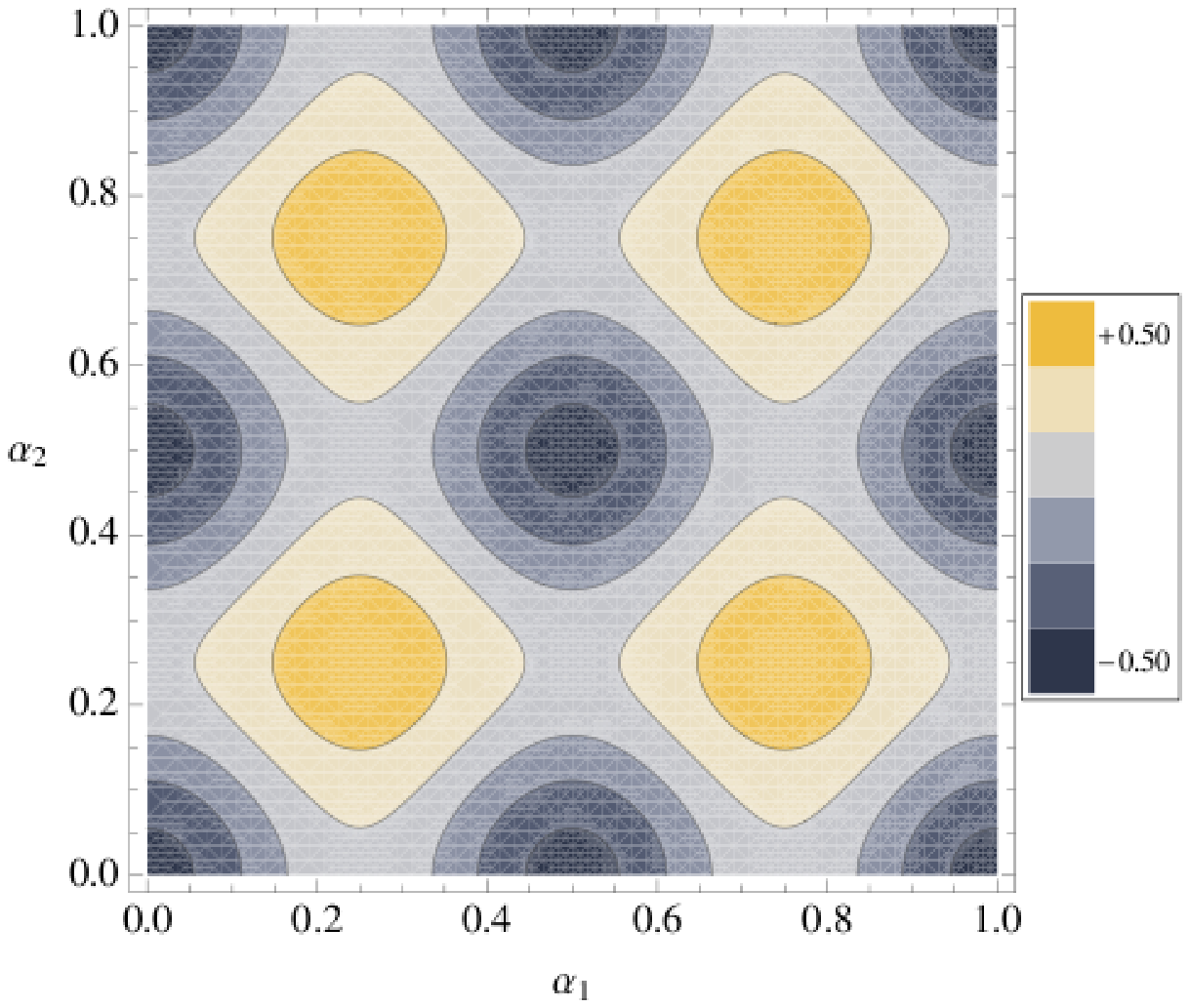,width=6.5cm} &
  \epsfig{file=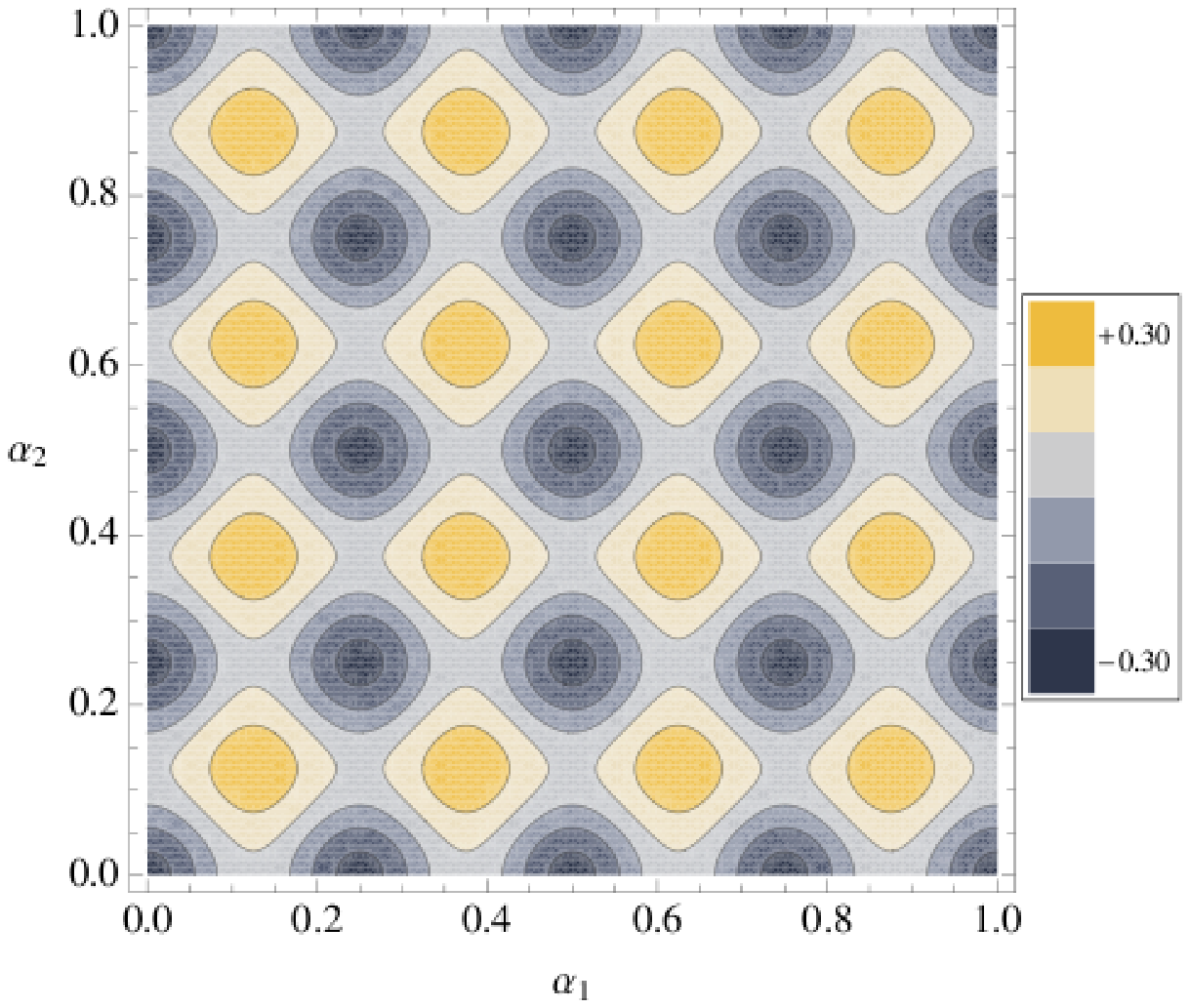,width=6.5cm}
\end{tabular}
\label{fig1}
\caption{Plot of $\Gamma_{(1)}^{q=1}$ for $\mathcal{K}=2$ and $\NK=2$ (left) and $\NK=3$ (right) 
as function of the SS phases $\alpha_1,\alpha_2$. Lighter (darker) regions indicate maximum (minimum) 
of the effective potential.}
\end{center}
\end{figure}

From Eq.~(\ref{effpotmatter}) one can see that the contribution from complex scalar fields in the 
adjoint representation comes with the same sign and a factor 1/2 compared to the gauge/ghost one. 
Consequently adding scalar matter fields does not affect the position of the minimum of the one-loop 
effective potential. Conversely, the contribution from fermions in the adjoint comes in Eq.~(\ref{effpotmatter})
with an opposite sign with respect to the contribution of the gauge/ghost fields. As a consequence, 
adding fermionic fields in the adjoint does not change the extrema of the theory, although it can turn 
maxima into minima and viceversa. The total effective potential in a model with $n_f$ $(6D)$ Weyl 
fermions and $n_s$ $(6D)$ complex scalar degree of freedom in the adjoint representation is simply 
given by:
\be
\Gamma_{adj} = (2 - 2 n_f + n_s) \, \Gamma^{q=1}_{(1)} \, .
\ee
The necessary condition for the inversion of the extrema is thus $n_f > 1 + n_s/2$.




Once the gauge symmetry group ($N$) and the symmetry group breaking pattern ($\NK$) have been fixed, 
there is still the possibility to modify the positions of the minima of the one-loop effective potential 
(and consequently the vevs of the dynamical symmetry breaking) by choosing conveniently the 
representation (weights) of matter fields. In the previous example we calculated the contribution to 
the one-loop effective potential using exclusively the adjoint representation (which has weights $q=\pm1$).
Let's consider now, instead, the contribution to the effective potential of a Weyl fermion belonging to 
the {\bf 5} representation of $SU(2)$. In this case we have both the contribution from weight 1 and weight 
2 fields. The one-loop effective potential for weight 2 fields in the {\bf 5} representation reads:
\bea 
\Gamma^{f,s}_{(1)}  & = & - \,2 \, \eta_{f,s} \, \NK^2 \, \frac{V^{4+2}}{\pi^{3}} \,\left\{\,2 \sum_{n_1,n_2=1} 
\frac{\cos (4\pi \NK n_1\alpha_1 )\,\cos (4 \pi \NK n_2\alpha_2 )}{\left[ (\NK l_1 n_1)^2 + (\NK l_2 n_2)^2\right]^{3}} 
\right. \nn \\
&+& \left.\sum_{n_1=1} \frac{\cos (4\pi \NK n_1 \alpha_1 )}{ (\NK l_1 n_1)^6} +\sum_{n_2=1} \frac{\cos (4\pi \NK n_2\alpha_2 )}{ (\NK l_2 n_2)^6} \right\} \, = \, \eta_{f,s} \, \Gamma^{q=2}_{(1)}
\label{vec_ghost_SU(K)_literature2}
\eea

\begin{figure}[!t]
\begin{center}
\begin{tabular}{cc}
\epsfig{file=GGH1b.ps,width=6.5cm} & 
\epsfig{file=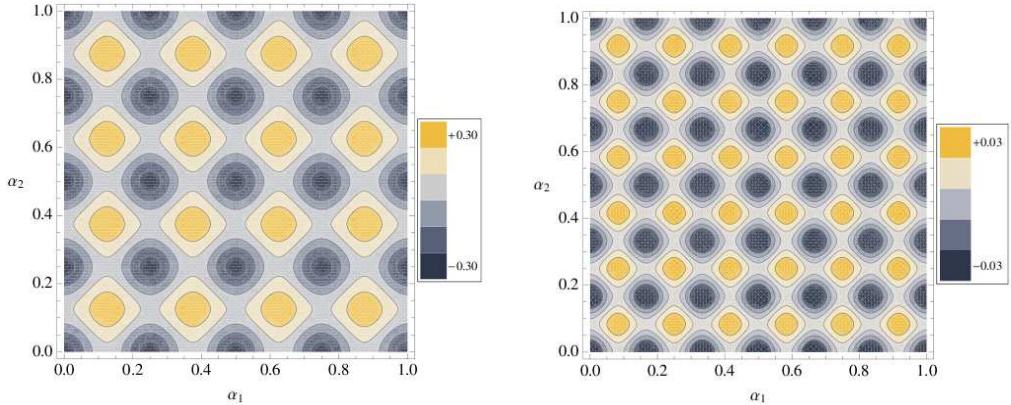,width=6.5cm} 
\end{tabular}
\label{fig2}
\caption{Plot of $\Gamma^{q=2}_{(1)}$ for $\mathcal{K}=2$ and $\NK=2$ (left) and $\NK=3$ (right) 
as function of the SS phases $\alpha_1,\alpha_2$. Lighter (darker) regions indicate maximum (minimum) 
of the effective potential.}
\end{center}
\end{figure}



For the sake of exemplification, let's consider a toy model with an original $SU(4)$ gauge symmetry 
broken down explicitly to $SU(2)$ by a $m=2$ 't Hooft flux and let's include the following matter fields: 
\begin{itemize}
\item One $6D$ Weyl fermion in the {\bf 5} representation of $SU(2)$;
\item One $6D$ complex scalar in the adjoint representation of $SU(2)$.
\end{itemize}
The one-loop effective potential for this field content is given then by
\be
V_{eff} = \frac{1}{V^{4+2}} \left( \Gamma_{(1)}^{g+gh} + \Gamma_{(1)}^f + \Gamma_{(1)}^s \right) \, = \, 
 \frac{1}{V^{4+2}} \left( \Gamma^{q=1}_{(1)} - 4 \, \Gamma^{q=2}_{(1)} \right) \, ,
\ee
where $\Gamma_{(1)}^s = \Gamma^{q=1}_{(1)}$ and $ \Gamma_{(1)}^f = -2\left( \Gamma^{q=1}_{(1)} 
+ 2\, \Gamma^{q=2}_{(1)} \right)$ to include both the weight 1 and weight 2 fields in the {\bf 5} 
representation of $SU(2)$. 
In Fig.~(\ref{fig2}) we plot the effective potential for this toy model assuming $l_1=l_2=l$ and setting the 
volume factor $\pi^3 l^6 =1$ for definiteness. As one can see in Fig.~(\ref{fig3}) the effective potential has 
a minimum for $\alpha_1=\alpha_2 = 0.1184$. For this value of the SS parameters $\alpha_i$ the $SU(2)$ 
symmetry is dynamically broken to $U(1)$ by the usual (rank preserving) Hosotani mechanism. We have 
provided in such a way a toy model where a double symmetry breaking has occurred. The first symmetry 
breaking is explicit and can be thought as the mechanism breaking the GUT symmetry to the SM gauge 
group, while the second dynamical (spontaneous) symmetry breaking could be seen as the EW symmetry 
breaking of the SM. An intrinsic problem of this mechamism is due to the fact that the two scales at which 
these breakings occur are connected to the same geometry factor $M \approx 1/l$. Nevertheless the 
spontaneous symmetry breaking depends explicitly on the weight of the fields in the $SU(2)$ representantion, 
while the 't Hooft breaking does not. So higher weights provide smaller values for the phases $\alpha_i$ and 
consequently a smaller value for the EW symmetry breaking scale. However, even if possible,  it seems to 
require some (unwanted) fine tuning to obtain in such a way a two-orders-of-magnitude separation between 
the scales of the two breakings. 

Of course this toy model is still far to represent a realistic pattern of the SM symmetry breaking. For example, 
following the previous results one could start with an $U(N)$ gauge theory broken by the 't Hooft flux to 
$SU(2)\times U(1)$ and subsequently to $U(1)\times U(1)$ by one-loop effects. Due to the fact that the 
Hosotani mechanism is a rank preserving breaking, in this toy model one would end with a massless $Z_0$ 
boson in the spectrum. A possible cure to this problem can be found introducing additional symmetry 
breaking mechanism as for example an orbifold structure\footnote{See for example the symmetry 
breaking patterns studied by \cite{vonGersdorff:2007uz}.}. 
A deeper and more complete study should be required in order to obtain a ``realistic'' GUT symmetry 
breaking model. Our interest in this paper was to point out in general the practical feasibility of the Hosotani 
mechanism in the presence of a non-vanishing 't Hooft flux. 
\begin{figure}[ht]
\begin{center}
\epsfig{file=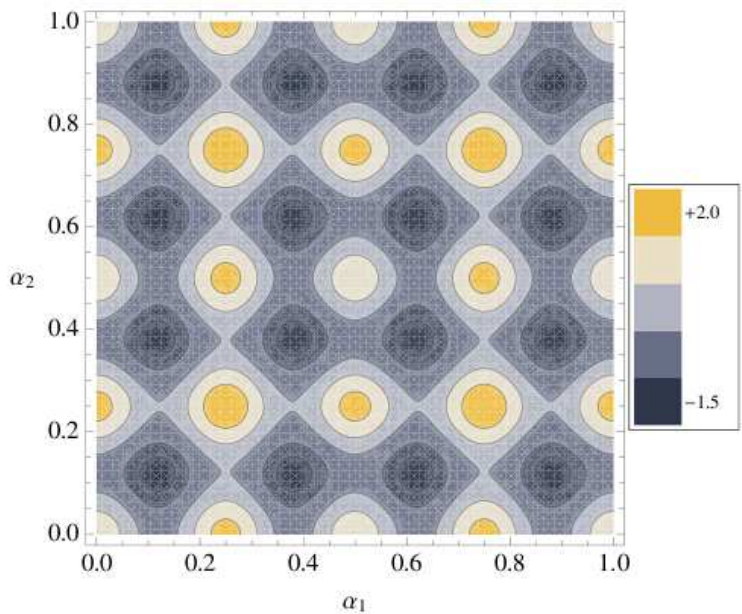,width=10cm}
\label{fig3}
\end{center}
\caption{\em The effective potential for the toy model discussed in the text as function of the SS phases 
$\alpha_1,\alpha_2$. Lighter (darker) regions indicate maximum (minimum) of the effective potential.}
\end{figure}

Apart from the gauge boson masses one should also calculate the one loop masses of the four dimensional 
scalars arising from the extradimensional components of the gauge fields. In particular one is interested in 
the scalars that correspond to the conserved symmetries and are given by the second derivatives of the 
effective potential with respect to $\alpha_1$ and $\alpha_2$. We discussed briefly this issue in 
subsection \ref{scalarsplit}.


%% file: conclu.tex
\section{Conclusions}
\label{conclusions}

The Hosotani mechanism is a very interesting symmetry breaking mechanism that arises in models 
defined in non simply-connected space-times, in which one has to specify the periodicity conditions 
of fields around the non-contractible cycles. It has been frequently applied in extra-dimensional model 
building to surrogate the SM electroweak symmetry breaking. While in five-dimensional models, 
$M_4 \times S^1$, the Hosotani mechanism completely describes the symmetry breaking pattern, in 
higher dimensional compactifications an additional ingredient has to be taken into account: the 't Hooft 
(non-abelian) flux. This flux appears as a consistency condition once we impose that the value of the 
gauge field  has to be independent of the path which has been followed to reach the starting point after 
wrapping the non-contractible loops, modulo a constant element of the center of the group. For this 
to be non-trivial one clearly needs at least two non-simply connected extra dimensions and thus we have 
focused in the case of a two-torus, that is $M_4 \times \mathcal{T}^2$. 

On the other hand, we have selected $U(N)$ as the gauge group for two phenomenological reasons. 
First, even when the 't Hooft flux is non-vanishing the theory admits the presence of fields in the 
fundamental representation. Secondly, since the 't Hooft flux is intimately related to the existence 
of a constant background magnetic flux for the $U(1) \subset U(N)$, it induces four-dimensional 
chirality for fundamental fermions through the usual mechanism \cite{Randjbar}. This is important 
because in the two-torus all stable $SU(N)$ background configurations are trivial \cite{Salvatori:2006pb} 
and therefore the non-abelian piece of the group could not do the job.

In this scenario, the symmetry breaking pattern for a $U(N)$ gauge theory strongly depends on an 
integer parameter $m=0,\dots,N-1$ (modulo N). For trivial values of the 't Hooft flux, $m=0$, one 
recovers the ``usual" Hosotani mechanism with two different non-integrable phases. This breaking 
is rank preserving because the Cartan subalgebra always remains unbroken. In the case of non-vanishing 
't Hooft flux, $m\neq 0$, two different processes occur simultaneously: an explicit symmetry breaking 
associated to the non-vanishing flux and a spontaneous and dynamical one, associated to the Hosotani 
mechanism. The explicit breaking due to the flux can reduce the rank of the group and thus has a 
different phenomenology than the previous one. 

In this paper we have, for the fist time, completely described the Hosotani mechanism in the presence 
of a non-trivial 't Hooft flux. In particular, we have calculated the mass spectrum both for the gauge 
fields and associated scalars and for fermions in different representations. Due to its sensitivity to 
the center of $U(N)$, the nature of the fermionic spectrum for the fundamental representation is 
peculiar. We have mentioned the possibility of obtaining chiral four-dimensional matter. The discussion 
of how fermions get masses and mix is, however, beyond the scope of this paper. 

A well known fact of the Hosotani mechanism is the degeneracy of the vacuum at tree level, and this is 
inherited in our model. A study of radiative corrections is therefore customary for obtaining both 
the true vacuum with the surviving symmetry and the values of the masses. With this aim, we have 
computed the one-loop effective potential for the general case of non-vanishing 't Hooft flux. We have 
found a very compact form in terms of the corresponding Wilson loops that can be particularized 
to the desired representation. Notice that for $m\neq 0$, matter in a representation sensitive to 
the center of the group does not help in removing the degeneracy since its contribution to the 
effective potential is a constant independent of the parameters that characterize the pattern of 
symmetry breaking.

We described a toy model to show explicitly how the mechanism work. We started with an $U(4)$ 
model broken down to $SU(2)\times U(1)$ by the 't Hooft flux and subsequently broken to $U(1) 
\times U(1)$ once a specific set of matter fields is chosen. We have provided in such a way a toy 
model where a double symmetry breaking has occurred, without having the need to introduce any 
additional structure. The first symmetry breaking is explicit and can be thought as the mechanism 
breaking the GUT symmetry to the SM gauge group, while the second dynamical (spontaneous) 
symmetry breaking can be seen as the EW symmetry breaking of the SM. Of course some extra 
work is needed in order to obtain a phenomenologically viable model. The deeper problem is obviously 
related to the presence of two massless gauge bosons once trying to reproduce the SM symmetry 
breaking. Some additional mechanism, like for example an orbifold structure, shoud be then advocated 
for giving mass to the $Z_0$. Netherveless it seems to us that the connection between the 't Hooft 
and the Hosotani mechanisms offers new and very interesting possibilities for model-builders.

%% file: appendix_new.tex
%
\section{Wave functions in the fundamental representation}
\label{AppendixA}
%

In this appendix we explicitly compute the wave function of a field, belonging to the $U(N)$ 
fundamental representation and living on a 2D torus with specific $U(N)$ periodicity conditions 
represented by the twists $T_a(y)$.


The general wave-function of a field in the fundamental representation living on a 2D torus 
with non-trivial periodicity condition is well known. Let's follow here the usual procedure 
and generalize it to the case of non-trivial 't Hooft flux. Let's indicate with $\Psi^{(p)} (y)$ 
the solution of the harmonic oscillator eigenvalue problem:
\bea
a^\dagger a \,\, \Psi^{(p)} (y) \,\,=\,\,p \, \Psi^{(p)} (y) \hspace*{2em} \mathrm{,} 
\hspace*{2em} p \,\in\, \mathbb{N} \,.
\label{eig_problem_app}
\eea
The creation and annihilation operators $a$ and $a^\dagger$ are defined in terms of the 
extra-dimensional components of the covariant derivatives $D_z$ and $D_{\bar{z}}$ as:
\bea
a = \sqrt{\frac{ N \mathcal{A}}{4 \pi m}} D_{\bar{z}} \quad  &,& \quad 
a^\dagger = -\sqrt{\frac{N \mathcal{A}}{4 \pi m}} D_{z} \,.
\label{creazione_distruzione}
\eea 
The wavefunction $\Psi^{(p)}(y)$ satisfies the following periodicity conditions
\bea
\Psi^{(p)} (y + \ell_a) \,\,\,= \,\,\, e^{\epsilon_{ab} i \pi \frac{m}{N} \frac{y_b}{l_b}}\,
\omega_a \,\,P^{s_a}\,\,Q^{t_a} \,\,\Psi^{(p)} (y) \,  
\label{perio_cond_app}
\eea
where we have expressed the general $U(N)$ twists in the {\em symmetric} gauge in terms 
of the 't Hooft matrices $P$ and $Q$, using the definitions in Eq.~(\ref{gaiarda}). The 
solutions of the problem of Eq.~(\ref{eig_problem_app}) with periodicity conditions of 
Eq.~(\ref{perio_cond_app}) constitute the generalized Landau levels. 
As in the standard harmonic oscillator case, it is possible to compute first the zero mode, satisfying
$a \,\Psi^{(0)} (y)=0$ and, subsequently, obtain all the higher modes by recursively applying the 
creation operator, $a^\dagger$. In the rest of the appendix we will uniquely concentrate in deriving 
the zero mode and consequently from now on we will drop the index $0$.

The wavefunction $\Psi (y) $ can be decomposed in $\widetilde{N}$ components:
\bea
\Psi (y) \equiv \left(\psi_1(y) \, , \,  \dots \, , \, \psi_j(y) \, , \, \dots \, , \psi_{\widetilde{N}} (y)\right)^T \, , \nn
\eea
with $\psi_j(y)$ $\mathcal{K}$-dimensional vectors of components:
\bea
\psi_j (y) \equiv \left(\psi_{j,1}(y) ,\, \dots ,\, \psi_{j,k}(y)  ,\, \dots \,,\psi_{j,\KK} (y) \right)^T \, .\nn 
\eea
%
The vectors $\Psi$ and $\psi_j$ can be viewed, in practice, as fundamental representations of, 
respectively, $U(\NK)$ and $U(\KK)$. Notice that while the operators $a$ and $a^\dagger$ act 
diagonally on the fundamental $U(\NK)$ representation 
\bea
a \,\Psi \,=\, 0 \quad \rightarrow \quad a \,\psi_{j} = 0 
\qquad \quad \forall \,j=1,..,\NK \, ,
\label{darisolvere}
\eea
the periodicity conditions in Eq.~(\ref{perio_cond_app}), instead, mix the $\NK$ components $\psi_{j}$ 
of $\Psi$ (while leaving unchanged the $\mathcal{K}$ components $\psi_{j,k}$ of $\psi_j$). 
In fact, Eq.~(\ref{perio_cond_app}) written in components of the $\NK$ representation reads\footnote{For 
definiteness, we will consider here the case $s_1=t_2=0$, $t_1=-1$ and $s_2=\mK$. Any other choice of 
the coefficients $s_a, t_a$ satisfying the constraint of Eq.~(\ref{st_constraint}) is of course equivalent.}: 
\bea
\psi_{j} (y + \ell_1) &=& e^{i \pi \frac{\mK}{\NK} \frac{y_2}{l_2}}\, \omega_1\, 
       e^{i \pi \left( \frac{1-\NK}{\NK}\right)} \,e^{2 \pi i \frac{1}{\NK}(j-1)} \,\psi_{j} (y) 
\label{perio_non_diag_app1} \\
\psi_{j} (y + \ell_2) &=& e^{-i \pi \frac{\mK}{\NK} \frac{y_1}{l_1}}\, \omega_2 \,
      e^{i \pi \mK \left( \frac{\NK-1}{\NK}\right)} \,\,\psi_{j + \mK} (y) \,.
\label{perio_non_diag_app2}
\eea 

The standard trick to diagonalize such periodicity conditions consists in repeating $\NK$ times the 
fundamental shift of length $l_a$. Introducing the following (diagonal) $\KK \times \KK$ phases matrices 
\bea
\begin{array}{lccc}
e^{2 \pi i \widehat{\gamma}_1}= e^{i \pi \left(1- \NK \right)} \, \omega_1^{\NK} \quad & , & \quad 
e^{2 \pi i \widehat{\gamma}_2}= e^{i \pi \mK \left(\NK-1\right)} \, \omega_2^{\NK} 
\end{array} 
\eea
and defining $L_a = \NK\,l_a$ and $d = \mK \NK$, the new periodicity conditions, for the $\KK$ 
dimensional vectors $\psi_j(y)$, read:
\bea
\psi_{j} (y + \NK \ell_a) \,=\, e^{i \pi d \epsilon_{ab} \frac{y_b}{L_b}}\,
    e^{2 \pi i \widehat{\gamma}_a} \psi_{j} (y) \,.
\label{perio_cond_diag_app}
\eea
Now, therefore, we want to find the harmonic oscillator zero mode with the periodicity conditions 
given in Eq.~(\ref{perio_cond_diag_app}). A possible ansatz for the wave function $\psi_{j}(y)$, 
compatible with the periodicity condition along  the direction $y_1$ is 
\bea
\psi_{j}(y) = \sum_{n=-\infty}^{\infty} \,\,e^{ i \pi d \frac{ y_1 y_2}{L_1 L_2}}  
\,\,e^{2 \pi i  \frac{y_1}{L_1} (n +\widehat{\gamma}_1)} \, \, C_{j,n} (y_2) \qquad 
\mathrm{for} ~j=1, \dots , \NK \;.
\label{prop+}
\eea
Here $C_{j,n} (y_2)$ are $\KK$ dimensional functions of the $y_2$ coordinate. To satisfy the 
periodicity condition along the direction $y_2$,  Eq.~(\ref{perio_cond_diag_app}) imposes 
that the coefficients $C_{j,n}(y_2)$ must satisfy the following condition:
\bea
C_{j,n} (y_2 + L_2) = e^{ 2 \pi i  \widehat{\gamma}_2} C_{j,n+ d} (y_2) \;.
\label{porca}
\eea
The explicit expression for the coefficients $C_{j,n} (y_2)$ is obtained substituting Eq.~(\ref{prop+}) 
in Eq.~(\ref{darisolvere}), that gives: 
\bea
\partial_2 C_{j,n} (y_2)=-\left(\frac{2\pi \,d}{L_1 L_2} y_2+\frac{2\pi}{L_1} (n+\widehat{\gamma}_1)\right) 
C_{j,n} (y_2) \,, \label{eqdif+}
\eea
with solution  
\bea
C_{j,n} (y_2) \,=\, e^{-\frac{\pi \,d }{L_1 L_2}  y_2^2} \, e^{- 2 \pi (n+\widehat{\gamma}_1)\frac{y_2}{L_1}} 
\, A_{j,n}\;.
\eea
The coefficient $A_{j,n}$ are then determined by the periodicity condition of Eq.~(\ref{porca}), implying
\bea
A_{j, n+d} \,=\, e^{-2\pi \frac{L_2}{L_1} \left(n+\widehat{\gamma}_1+d/2 \right)} \,e^{-2 \pi i \widehat{\gamma}_2} 
\, A_{j,n} \,,
\eea
whose solution is 
\bea
A_{j,n} \,=\, e^{-\frac{\pi}{d} \frac{L_2}{L_1} n^2} \,
                  e^{-2\pi i \left(\widehat{\gamma}_2 - i \frac{L_2}{L_1}\widehat{\gamma}_1 \right) \frac{n}{d}} \, B_{j,n}\, ,
\eea
with the constants $B_{j,n}$ satisfying the condition $B_{j,n+d} = B_{j,n}$. There exist, therefore, 
only $d$ arbitrary constant coefficients for each value of the index $j$ and, consequently, $d$ 
independent solutions for the zero mode of each component $\psi_{j}$. We will characterize them 
by the integer number $q = 0, ...,  d-1 $. All in all, the lightest wave function $j^\mathrm{th}$ component 
can be written as 
\bea
\psi_{j}(y) &=& \sum_{q=0}^{d-1} \, f_{q}(y) \, B_{j,q} \,,
\label{sol_deg_d}
\eea
where $B_{j,q}$ are, for each $j$, $d$ arbitrary ($\KK$ dimensional vector) coefficients subject to the 
normalization condition 
\bea
\sum_{q=0}^{d-1} | B_{j,q} |^2 =1\,,
\label{normb}
\eea
and $f_q(y)$ are the $d$ independent ($\KK \times \KK$ matrix) eigenfunctions given by 
\bea
f_{q}(y) & = & \left(\frac{2 d}{L_1^3 \,L_2}\right)^{\frac{1}{4}}
   e^{\frac{\pi i d}{L_1 L_2} y_2\left(y_1+i y_2\right)}\, 
   e^{\frac{2\pi i\widehat{\gamma}_1}{L_1}\left(y_1+i y_2\right)} \times \nn \\ 
  & & \hspace{1.15cm} \sum_{n=-\infty}^{\infty} e^{-\pi d\frac{L_2}{L_1} \left(n +q/d\right)^2}  
   e^{-2\pi i \left(\widehat{\gamma}_2-i\frac{L_2}{L_1}\widehat{\gamma}_1-
   \frac{(y_1+i y_2) d}{L_2}\right)\left(n +q/d \right)}\;.
\label{wavefunctionzero+}
\eea
Notice that the solutions $f_q(y)$ do not depend explicitly, at this stage, on the index $j=1,...,\NK$, 
while they depend, implicitly on the index $k=1,...,\mathcal{K}$ trough the phase matrices $\widehat{\gamma}_a$ 
that are diagonal (but in general not proportional to the identity) $\KK \times \KK$ matrix.  
The results in Eqs.~(\ref{sol_deg_d},\ref{normb},\ref{wavefunctionzero+}) express the general 
zero-mode solution of the generalized Landau problem on the $\NK l_1 \times \NK l_2$ torus with 
diagonal periodicity condition of Eq.~(\ref{perio_cond_diag_app}). We must now work backwards 
to recover the solution on the original $l_1 \times l_2$ torus. 

It is straightforward from Eq.~(\ref{wavefunctionzero+}) to check that the functions $f_{q}(y)$ 
satisfy the following periodicity conditions under the fundamental shifts $l_1, l_2$:
\bea
f_q (y+\ell_1) &=& e^{i \pi \frac{\mK}{\NK} \frac{y_2}{l_2}} \, \omega_1\,
    e^{i \pi \left( \frac{1-\NK}{\NK}\right)}\,e^{2 \pi i \frac{q}{\NK}} \,\,f_{q} (y) 
\label{prop_f_non_diag_app1} \\
f_q (y+\ell_2) &=& e^{-i \pi \frac{\mK}{\NK}\frac{y_1}{l_1}} \, \omega_2\,
    e^{i \pi \mK \left( \frac{\NK-1}{\NK}\right)} \,\, f_{q + \mK} (y) \,.
\label{prop_f_non_diag_app2}
\eea  
Substituting Eq.~(\ref{sol_deg_d}) in the periodicity conditions of Eqs.~(\ref{perio_non_diag_app1},
\ref{perio_non_diag_app2}) and using the properties of Eqs.~(\ref{prop_f_non_diag_app1},
\ref{prop_f_non_diag_app2}), it is possible to verify that the solution is consistent only 
if the following two conditions are satisfied:
\bea
   \sum_{q=0}^{d-1} e^{2\pi i \frac{q}{\NK}} \,f_{q} \, B_{j,q} & = & 
   e^{2\pi i \frac{j-1}{\NK}} \sum_{q=0}^{d-1} \,f_{q} \, B_{j,q} \label{cond1} \\
   \sum_{q=0}^{d-1} \,f_{q} \, B_{j+\mK,q} & = & \sum_{q=0}^{d-1} \, f_{q+\mK} \, B_{j,q} \,. 
   \label{cond2}
\eea
The condition, Eq.~(\ref{cond1}), is satisfied only if $q = q' \NK +j-1$, with $q' = 0,1, \dots , \mK-1$. 
So, as expected, in the original torus there are only $\mK$ independent ($\KK$ dimensional) solutions, 
instead of the $d=\mK \NK$ ones that are allowed in the extended torus. Using Eq.~(\ref{cond1}), and 
the facts that $q = q+d$ and $\NK/\mK$ cannot be an integer, one obtains that Eq.~(\ref{cond2}) is 
satisfied only if $B_{j,q}=B_{q}$, i.e. the $B_q$ are j-independent constant ($\KK$ dimensional) coefficients..
One can imagine this reduction operates in the following way. First, for each one of the $N$ directions of 
the $SU(N)$ fundamental it divides by $\NK$ 
the number of independent degrees of freedom. Secondly, the $N$ components of the fermion multiplet in 
the $SU(N)$ fundamental are gathered in $\mcl{K}$ sets of $\NK$ fermions. By doing this one finds 
that only one independent degree of freedom remains for each of these sets of $\NK$ fermions. 

Finally, the zero-mode solution of the eigenvalue problem in Eq.~(\ref{eig_problem_app}) with the periodicity 
conditions in Eq.~(\ref{perio_non_diag_app1},\ref{perio_non_diag_app2}) is given by:
\bea
\Psi^{(0)} (y) \,=\, \left(\psi^{(0)}_1(y) \, , \,  \dots \, , \, \psi^{(0)}_j(y) \, , \, \dots \, , 
      \psi^{(0)}_{\widetilde{N}} (y)\right)^T \, , \nn
\eea
with 
\bea
\psi^{(0)}_j (y) \, = \, \sum_{q=0}^{\mK-1} \,f_{q\NK +j -1}(y) \, B_{q} \nn 
\eea
$\mathcal{K}$-dimensional vectors linear combination of the $\mK$ independent functions $f_q(y)$ 
($\KK \times \KK$ diagonal matrices) which general expression is written in Eq.~(\ref{wavefunctionzero+}) 
and $\mK$ independent coefficients $B_q$ ($\KK$-dimensional vectors). Therefore there are in total $m$ 
degrees of freedom. Notice that the explicit symmetry 
breaking $SU(N) \rightarrow SU(\KK)$ due to the 't Hooft flux is made explicit through the $j$-index 
dependence of the wavefunctions $f_{q\NK +j-1}(y)$, that localize the solutions at different points 
of the torus. In the case in which all $SU(\mathcal{K})$ continuous phases $\alpha_a$ are zero, these degrees 
of freedom form $\mK$ independent fundamental representations of $U(\mathcal{K})$: in this case indeed 
$(f_q)_{11}= (f_q)_{22}= ... = (f_q)_{\KK \KK}$. On the contrary, for non trivial phases $\alpha_a$, 
different entries of the fundamental $U(\NK)$ representation may have different wave function. Notice 
that the $U(\mathcal{K})$ breaking  manifests itself only in the form of wavefunction: the eigenvalues of the 
number operator $a^\dagger a$ (and consequently the effective $4D$ masses) are completely determined 
by the commutation rules in Eq.~(\ref{comm_rules_app}) and they do not depend on the $SU(\mathcal{K})$ 
continuous phases.

%% file: appendix_heat.tex
\section{The heat kernel and the effective action: the computation}
\label{AppendixB}

The heat kernel is a very efficient way of calculating quantum effects in field theories 
defined on general manifolds\footnote{We will consider here only the flat manifold case, but 
all the formalism can be easily extended to curved ones. See for example \cite{Vassilevich:2003xt}
for an extensive review on the subject.}. The reason relies in its intimate connection with the one-loop 
effective action, explicitly
\begin{equation}
\Gamma_{(1)} = \frac{1}{2}\log\det\,\Delta=\frac{1}{2}\trm{Tr}\log\,\Delta=-
\frac{1}{2}\int_0^\infty\frac{dt}{t}\,G(t) \,.
\label{starting}
\end{equation}
Here $\Delta$ is the operator in the quadratic part of the action, usually resulting from the expansion 
around an arbitrary background field, and $G(t)$ the kernel of $\Delta$. Notice that it contains a trace 
over the adequate discrete indices (Lorentz, gauge...). The kernel $G(t)$ can be rewritten as 
\begin{equation}
G(t)=\int d^{4}x\,\int d^{d}y\,\,\,G\left(\{x,y\},\{x,y\},t\right) \,,
\end{equation}
in terms of a heat function, $G\left(\{x_i,y_i\},\{x_f,y_f\},t\right)$, that satisfies the heat equation
\begin{equation}
 \Delta_{\{x_i,y_i\}} \, G\left(\{x_i,y_i\},\{x_f,y_f\},t\right) = - \frac{\partial}{\partial t} \,G\left(\{x_i,y_i\},\{x_f,y_f\},t\right) \,,
\label{heat_eq}
\end{equation}
with initial condition
\begin{eqnarray}
G\left(\{x_i,y_i\},\{x_f,y_f\},t=0\right) \,=\,\delta^{4} (x_i-x_f)\,\delta^{ED}(y_i - y_f).
\label{in_con}
\end{eqnarray} 
In the previous equation by $\delta^{ED}$ we mean the appropriate delta function defined in 
the specific extra-dimensional manifold. In terms of the eigenfunctions, $g_n$, and the eigenvalues, 
$\lambda_n$, of the bilinear operator $\Delta$, the heat function takes the form
\begin{equation}
G\left(\{x_i,y_i\},\{x_f,y_f\},t\right) \equiv \sum_n e^{-\lambda_n t} \,g_n(\{x_f,y_f\}) \,g_n^*(\{x_i,y_i\}) \,.
\label{G_ff*}
\end{equation} 
Here the eigenvalues are assumed positive, real and discrete, which will be the case in what follows. 
The initial condition of Eq.~(\ref{in_con}) results as a straightforward consequence of the 
eigenfunctions completeness relation.

The effective action is in general a divergent quantity and requires regularization.
A very elegant way of doing so is using $\zeta$-function techniques. The generalized 
$\zeta$-function associated to the operator $\Delta$ is defined by
\begin{equation}
\zeta_\Delta(s) \,=\, \sum_{n} \,\frac{1}{\lambda_n^s} \,,
\end{equation}
and it is related to the heat kernel by a Mellin transformation
\begin{equation}
\zeta_\Delta (s) \,= \, \frac{1}{\Gamma(s)}\, \int_0^{\infty} dt \,t^{s-1} \,\,G(t) \,,
\label{mellin}
\end{equation}
in such a way that the one-loop effective action is simply
\begin{equation}
\Gamma_{(1)} = -\frac{1}{2}\zeta_\Delta^{\prime}(0) \, .
\label{gammareg}
\end{equation}
The regularization of $\Gamma_{(1)}$ is provided through analytic continuation 
\cite{Dowker:1975tf,Hawking:1976ja} to 
\begin{equation}
\Gamma_{(1)}(s,\mu)=-\frac{1}{2} \,\mu^{2s}\, \Gamma (s)\,\zeta_\Delta (s) \, ,
\end{equation}
being $\mu$ an appropriate regularization scale. In the limit $s \to 0$ one obtains the ($\overline{\mathrm{MS}}$) renormalized effective action:
\begin{equation}
\Gamma^{ren}_{(1)}(\mu)=-\frac{1}{2} \,\zeta_\Delta'(0) - \frac{1}{2} \log \mu^2 \zeta_\Delta (0)~.
\label{accion_renormalizada}
\end{equation}

Our computational strategy will be thus to solve the heat equation (\ref{heat_eq}) with the relevant initial condition, insert the solution in (\ref{mellin}) and get the renormalized effective action trough the $\zeta$-function. Calculating the heat function instead of the heat kernel will be necessary to capture the non-local nature of the contributions we are looking for. 

All previous reasonings apply independently of the considered manifold. Now, suppose that the $y_a$ coordinates describe and extra-dimensional compact manifold. Then, at the level of the action, it is possible to expand the fields in harmonics of this manifold to get a four-dimensional theory with an infinite number of modes. Each of these KK modes has its own quadratic operator, for example in our case
\be\label{opermode}
\Delta_n=-\partial_\mu\partial^\mu+M_n^2
\ee  
where $M_n^2$ are the eigenvalues of the operator acting on the extra-dimensional coordinates. This term is perceived in four dimensions as a mass, different for each mode. It is natural then to compute the contribution to the effective potential, $\Gamma_{(1)}^n$, due to a single mode and associated to (\ref{opermode}) and simply add up the infinite tower, hoping that
\be\label{hope}
\Gamma_{(1)}=\sum_n\,\Gamma_{(1)}^n.
\ee
For a finite number of fields, this relation is safe. Unfortunately, the case of an infinite number of modes is much more delicate. For instance, it has been observed several times \cite{Alvarez:2006we} that in general the UV divergences and counterterms computed in the complete manifold $\delta\Gamma_{(1)}$ do not coincide with the ones obtained after summing the counterterms due to each particular mode, i.e.,
\be
\delta\Gamma_{(1)}\ne\sum_n\,\delta\Gamma_{(1)}^n.
\ee
In this respect, we are not aware of precise statements about finite or non-local contributions to the effective action. Having this in mind, we will perform the computation according to the two prescriptions implicit in (\ref{hope}). Let us start with the right hand side, that is, solving the heat equation for an operator of the form (\ref{opermode}) with the initial condition
\be
G(x_i,x_f,t=0) = \delta^4 (x_i - x_f). 
\ee
The form of the heat function in this case is well known to be
\be
G(x_i, x_f, t) \,= \,\frac{1}{(4 \pi t)^2}  e^{\frac{-(x_i-x_f)^2}{4 t}}\,e^{- M^2_n t  } \,,
\label{heat_funct_4d}
\ee
from which the regularized $\zeta$-function and the one-loop effective action read, respectively,
\begin{eqnarray}
\zeta_\Delta^R(s) \, &=&\,\frac{V^{4}}{(4 \pi)^2 } \,\,(M^2_n)^{2-s} \frac{\Gamma (s-2)}{\Gamma(s)}\,. \\
\Gamma_{(1)\mathrm{ren}}^{n}  &=& - \,\frac{V^{4}}{(4 \pi)^{2}} \,\,\left(M^2_n\right)^{2} 
\left( \frac{3}{4} - \frac{1}{2} \log \frac{M^2_n}{\mu^2} \right) \,.
\label{v_eff_4d}
\end{eqnarray}
Up to this point, we have not particularized the form of the spectrum $M_n^2$, but we must in order 
to evaluate the infinite sum. However, it is easy to check that the non-local (and finite) 
contribution to the one-loop effective action comes only from $6D$ fields which have vanishing covariant 
derivatives commutator and therefore are insensitive to the 't Hooft flux. On the contrary, fields in 
representations with a non-vanishing commutator give only a divergent constant, independent of the 
symmetry breaking parameters and irrelevant for determining the true vacuum. This should be clear from 
the absence of SS phases in the spectrum of fermions in the fundamental representation (\ref{ferspec}).

Consequently, in the following we will only concentrate on the first type of $4D$ degrees of freedom. 
For such $4D$ fields, the tree-level square-mass reads 
\be
M^2_{n(k)} \,\,=\,\,4 \pi^2\, \sum_{a=1}^2 \,\left( n_a \,+\,w_a^{(k)}\right)^2 \,\frac{1}{l^2_a} \,,
\label{M^2_k}
\ee 
where $(k)$ is a representation index and $w_a^{(k)}$ contains all continuous parameters characterizing 
the $U(N)$ vacua and appearing in the periodicity conditions and/or in the background (if we are not in the 
``symmetric gauge''). They are related to Wilson loops winding once the two non-contractible cycle of the 
torus as follows
\be
\left[W_a (y,y)\right]_{ik} \,=\,\left(\,\mathcal{P} e^{i g \int_{y}^{y+l_a} B_b dy^b}\, 
     T_a\,\right)_{ik} \,\equiv\,e^{2 \pi i w_a^{(k)}}\,\delta_{ik} \,.
\ee
Summing the effective potential (\ref{v_eff_4d}) for each four-dimensional mode of the form (\ref{M^2_k}) 
we are led to the evaluation of the following two series  
\bea
&1.&\hspace{1.5ex}\sum_n\,\Gamma_{(1)\mathrm{ren}}^{n}\,\supset\, \sum_{n_1,n_2} \left(\sum_{a=1}^2 \left(n_a + w_a \right)^2 \,\frac{4 \pi^2}{l_a^2} \right)^2\\
&2.&  \hspace{1.5ex}\sum_n\,\Gamma_{(1)\mathrm{ren}}^{n}\,\supset\,\sum_{n_1,n_2}\left( \sum_{a=1}^2 \left(n_a + w_a \right)^2 \,\,\frac{4 \pi^2}{l_a^2}\right)^2\,
         \log \sum_{a=1}^2 \frac{4 \pi^2 \left(n_a + w_a \right)^2}{l_a^2 \mu^2}
\eea
For the sake of simplicity in the previous equations and in the following lines we drop the index $(k)$ 
from the formulas. The first series may be computed as follows
\begin{eqnarray}
&&\sum_{n_1,n_2} \left(\sum_{a=1}^2 \left(n_a + w_a \right)^2 \,\frac{4 \pi^2}{l_a^2} \right)^2 =
\, \left.\frac{\partial^2}{\partial \xi^2}  \prod_{a=1}^2 \left( \sum_{n_a}e^{- \frac{4 \pi^2\xi}{l_a^2}\, 
           \left(n_a + w_a \right)^2 } \right) \right|_{\xi=0}\\
&&  \hspace{7ex}  = \, V^2\left.\frac{\partial^2}{\partial \xi^2} \frac{1}{(4 \pi \xi)} 
            \left( \sum_{m_1,m_2} e^{- \sum_{a=1}^2\,\frac{\left(l_a m_a \right)^2 }{4 \xi}} e^{2 \pi i w_a m_a}\right) 
            \right|_{\xi=0} \nn \\
&&  \hspace{7ex}  = \,2  \frac{V^2}{(4 \pi)} \sum_{m_1,m_2} \delta_{m_1,0} \,\delta_{m_2,0}  
           \left. \frac{1}{  \xi^{3}}\right|_{\xi=0}   e^{2 \pi i  \,\sum_{a=1}^2\,w_a m_a} \, = \, \frac{V^2}{ 2 \pi} 
\left. \frac{1}{ \xi^{\frac{4+d}{2}}}\right|_{\xi=0} \, . \nn
\label{serie_1}
\end{eqnarray}
We see that the first contribution to the $4D$ effective potential is independent of the continuous 
parameters appearing in the background and in the periodicity conditions. It gives rise to a divergence 
proportional to the volume. The calculation of the second series proceeds in a similar way:
\begin{eqnarray}
\sum_{n_1,n_2} && \left( \sum_{a=1}^2 \left(n_a + w_a \right)^2 \,\,\frac{4 \pi^2}{l_a^2}\right)^2\,\,
                    \log \sum_{a=1}^2 \frac{4 \pi^2 \left(n_a + w_a \right)^2}{l_a^2 \mu^2} = \nn \\
& &  =  - \int_0^\infty \frac{dt}{t} \,\frac{\partial^2}{\partial t^2}\,\prod_{a=1}^2 
     \left( \frac{l_a}{(4 \pi t)^{\frac{1}{2}}}\,\sum_{m_a} \,e^{- \frac{(l_a m_a)^2}{4 t}} \,e^{2 \pi i w_a m_a}\right) \nn \\
& &  = - \frac{V^2}{4 \pi}  \int_0^\infty \frac{dt}{t} \,\frac{\partial^2}{\partial t^2}\, \frac{1}{t}\,
     \sum_{m_1,m_2} \,e^{- \sum_{a=1}^2 \frac{(l_a m_a)^2}{4 t}} \,e^{2 \pi i \sum_{a=1}^2 w_a m_a}  \nn \\
& &  = - \frac{V^2}{4 \pi} \left[ 2\,\int_0^\infty \frac{dt}{t^{3}} \, + \,\sum_{m_1,m_2\neq 0}\,e^{2 \pi i  
     \sum_{a=1}^2 w_a m_a} \int_0^\infty \frac{dt}{t}  \frac{\partial^2}{\partial t^2}\, 
     \frac{1}{t}\,e^{-\sum_{a=1}^2 \frac{(l_a m_a)^2}{4 t}}  \right] \nn \\
& &  = - \frac{V^2}{ 2 \pi} \int_0^\infty \frac{dt}{t^{3}} - \,\frac{64 V^2}{ \pi}  \,
     \sum_{m_1,m_2\neq 0}\,\frac{W_1^{m_1}\,W_2^{m_2}}{\left[(l_1 m_1)^2 + (l_2 m_2)^2\right]^{3}} \,  .
\label{serie_2}
\end{eqnarray}
The first term in the last line is the divergent contribution from the zero mode, and consequently is proportional to the volume but independent of the continuous parameters characterizing the $U(N)$ vacua. The second term is the finite contribution we are interested in.

Using the results of Eqs.~(\ref{serie_1})-(\ref{serie_2}) and obviating the parameter-independent terms, 
the contribution to the one-loop effective action due to a $6D$ degree of freedom with $4D$ spectrum 
$M^2_{n(k)}$ of the form Eq.~(\ref{M^2_k}) is
\begin{eqnarray}
\left( \Gamma_{(1)}^{\mathrm{ren}} \right)_k \,=\,- \frac{V^{4+2}}{\pi^{3}}\, 
            \sum_{m_1,m_2\neq0}  \frac{\trm{Tr}\left(W_1^{m_1}\,W_2^{m_2} \right)} 
           {\left[(l_1 m_1)^2 +(l_2 m_2)^2\right]^{3}}  \,.
\end{eqnarray}
Particularizing the trace to the desired representation of both Lorentz and gauge group indices one gets 
the effective potential used in the main body of the paper. 

As a final check, we will repeat the computation but without any reference to the spectrum of the reduced theory, that is, solving directly the heat equation in $6D$. As we have mentioned, trapping non-local physics with the heat kernel is not an easy task. For this reason, we will consider only the more tractable case of vanishing 't Hooft flux, where a ``symmetric" gauge is fully accessible. In this particular gauge, the content of the theory is completely displaced to the non-trivial constant periodicity conditions while the background field can be switched off.

Our path to obtain the relevant contributions will be to reflect the desired periodicity in the initial conditions (\ref{in_con}). For another attempt along similar lines see \cite{vonGersdorff:2008df}. Consider the following ansatz for the extra-dimensional delta 
\be
\delta^{\mathcal{T}_2}(y_f-y_i)  \equiv  \sum_{m_a = -\infty}^{\infty} 
\delta^2 (y_f - y_i + m \cdot \ell)\,\,\,T_1^{m_1}\, T_2^{m_2} 
\label{delta_extra-dim_1}
\ee
where we use $m\cdot \ell$ as the short-hand notation for the coordinate shift $m_1 \ell_1 + m_2 \ell_2$.
The extra-dimensional coordinates, $y_{(i,f)}$, are defined in the fundamental domain of 
the torus, $y \in \left[0, l_a \right)$. The $\delta^2$ appearing on the right-hand side is the usual Dirac delta defined in the covering space ${\mathcal R}^2$. The integers $m_a$ are the winding numbers that account for how many times one has to wind around the cycle ``\textit{a}'' in order to connect the coordinates $y_i$ and $y_i +  w\cdot \ell$ in the covering space.
One gets a factor of the twist for each of these windings. Their presence in the initial condition ensures the desired periodicity of the heat function and therefore of the effective potential, as well as their gauge invariance. Note that this expression makes sense since the twists are point-independent and commute in the absence of flux\footnote{This ansatz is inspired in studies of the heat kernel in finite temperature field theories, in which Euclidean time is compactified into a circle. The heat function can be expressed as an infinite sum of zero temperature (that is, uncompactified) heat kernels as shown in \cite{Dowker:1976pr}. Our initial condition is just a generalization to non-trivial twists.}.

Now we are in a position to solve the heat equation with the previous initial condition. Let us consider the contribution to the one-loop effective potential due to a field in a generic representation $\mathcal{R}$ of $U(N)$. In the symmetric gauge the operator is a flat Laplacian so the heat function is again easily guessed
\bea
G\left(\{x_i,y_i\},\{x_f,y_f\},t\right) \,=\, \sum_{m_1,m_2}\,\frac{1}{(4 \pi t)^{3}} \, e^{-\frac{1}{4 t} 
 \left[ (x_f-x_i )^2 +  (y_f-y_i + m \cdot \ell)^2\right]}\,\,T_1^{m_1}\, T_2^{m_2}  
\label{heat_sol}
\eea
The overall constant factor has been fixed using the definition of the  Dirac delta:
\begin{eqnarray}
\delta (x) \,\equiv\, 
\lim_{\epsilon \rightarrow 0^+} \, \frac{1}{\sqrt{4 \pi \epsilon}}\,\,\,\, e^{- \frac{x^2}{4 \epsilon}} \,.
\end{eqnarray} 
From this solution, the associated $\zeta$-function is
\bea
\label{zeta_R_i} \hspace{-6ex}
\zeta_\Delta(s) \,&=&\frac{V^{4+2}}{(4 \pi)^{3} \Gamma (s)} \left[\left.\frac{t^{s-3}}{s-4}\right|_{t=0}^{t=\infty}\hspace{-0.2cm} + \hspace{-0.1cm} 
\sum_{m_1,m_2\neq0} \hspace{-0.3cm} \mathrm{Tr} \Big( W_1^{m_1}\,W_2^{m_2} \Big) \hspace{-0.15cm} 
\int_0^{\infty} \hspace{-0.35cm} dt\,t^{s-4}  e^{-\frac{1}{4 t} \sum_{a=1}^2 \left( l_a m_a \right)^2} \right] 
\nn
\eea
where $V^{4+2}$ is the $6D$ volume, $\mathrm{Tr}$ denotes the trace over the chosen $U(N)$ representation and we have used (\ref{Wloopsym}) to write the Wilson loop.
 
The first term in Eq.~(\ref{zeta_R_i}) comes from the $m_1=m_2=0$ contribution and it is divergent. The zero winding numbers case corresponds, in fact, to local operator contributions and it is independent of the continuous $U(N)$ SS parameters. For $m_1$ and/or $m_2$ different from zero, the integral and the sum in the second term converge and so they can be safely interchanged. This contribution, in fact, proceeds from the Wilson loops that wrap around the non-contractible cycles of the torus at least once. 
 
The regularized $\zeta$-function finally reads:
\begin{eqnarray}
\zeta_A (s) = \,\frac{V^{4+2}}{\pi^{3}} \,\frac{\Gamma (3-s)}{4^s \Gamma(s)}
\sum_{w_1,w_2\neq0}\,\frac{\mathrm{Tr} \,\left(\,W_1^{w_1}\,W_2^{w_2}\,\right)} 
{\left[ (l_1 w_1)^2 + (l_2 w_2)^2\right]^{3-s}}  \,.
\end{eqnarray}
and consequently the effective action is given by:
\begin{eqnarray}
\Gamma_{(1)}^{\mathrm{ren}} &=& -\,\frac{V^{4+2}}{\pi^{3}}\,\sum_{w_1,w_2\neq0}\,
  \frac{\mathrm{Tr} \, \left(\,W_1^{w_1}\,W_2^{w_2} \,\right)}
  {\left[ (l_1 w_1)^2 + (l_2 w_2)^2\right]^{3}} \,.
\label{Gamma_1^ren_extra}
\end{eqnarray}
A comparison with the previous result obtained from the $4D$ spectrum shows immediately that the 
higher-dimensional and dimensionally reduced computations of the finite part of the effective action 
actually agree. Notice that this is not in contradiction with the statements of \cite{Alvarez:2006we} 
since there non-local sectors were not considered. Conversely, here we have discarded the local UV 
divergent contributions studied in those works.

%% file: eff_pot.bbl
\begin{thebibliography}{9}

\bibitem{lep}
See the LEP Electroweak Working Group WEB page for the latest fits on SM Higgs mass: 
http://lepewwg.web.cern.ch/LEPEWWG/

\bibitem{technicolour}
 L.~Susskind,
 Phys.\ Rev.\ D {\bf 20}, 2619 (1979).
%
\bibitem{littlehiggs}
 C.~T.~Hill, S.~Pokorski and J.~Wang,
 Phys.\ Rev.\ D {\bf 64}, 105005 (2001);
 N.~Arkani-Hamed, A.~G.~Cohen and H.~Georgi,
 Phys.\ Lett.\ B {\bf 513}, 232 (2001);
%
\bibitem{Manton}
 D.~B.~Fairlie,
 Phys.\ Lett.\ B {\bf 82}, 97 (1979) and
 J.\ Phys.\ G {\bf 5}, L55 (1979);
 N.~S.~Manton,
 Nucl.\ Phys.\ B {\bf 158}, 141 (1979);
 P.~Forgacs, N.~S.~Manton,
 Commun.\ Math.\ Phys.\  {\bf 72}, 15 (1980);
%
\bibitem{Casas}
  J.~A.~Casas, J.~R.~Espinosa and I.~Hidalgo,
  JHEP {\bf 0411} (2004) 057; 
  J.~A.~Casas, J.~R.~Espinosa and I.~Hidalgo,
  JHEP {\bf 0503} (2005) 038 
%
\bibitem{Alfaro:2006is}
 J.~Alfaro, A.~Broncano, M.~B.~Gavela, S.~Rigolin and M.~Salvatori,
 JHEP {\bf 0701}, 005 (2007).
%
\bibitem{Nielsen}
 N.~K.~Nielsen and P.~Olesen, 
 Nucl.\ Phys.\ B {\bf 144}, 376 (1978);
 N.~K.~Nielsen and P.~Olesen, 
 Phys.\ Lett.\ B {\bf 79}, 304 (1978);
 J.~Ambjorn, N.~K.~Nielsen and P.~Olesen, 
 Nucl.\ Phys.\ B {\bf 152}, 75 (1979);
%
\bibitem{Luscher:1982ma}
  M.~Luscher,
  Nucl.\ Phys.\  B {\bf 219} (1983) 233.
%
\bibitem{Hosotani}
Y.~Hosotani,
  Phys.\ Lett.\  B {\bf 126}, 309 (1983);
  Y.~Hosotani,
  Phys.\ Lett.\  B {\bf 129}, 193 (1983);
Y.~Hosotani,
  Annals Phys.\  {\bf 190}, 233 (1989);
Y.~Hosotani,
  arXiv:hep-ph/0408012;
Y.~Hosotani,
  arXiv:hep-ph/0504272.
%
\bibitem{Hetrick:1989jk}
  J.~E.~Hetrick and C.~L.~Ho,
  Phys.\ Rev.\  D {\bf 40}, 4085 (1989).
%
\bibitem{Mclachlan:1989pf}
  A.~T.~Davies and A.~McLachlan,
  Phys.\ Lett.\  B {\bf 200} (1988) 305.
  A.~Mclachlan,
  Phys.\ Lett.\  B {\bf 222}, 372 (1989)  [Erratum-ibid.\  B {\bf 237}, 650 (1990)];
  A.~McLachlan,
  Nucl.\ Phys.\  B {\bf 338}, 188 (1990).
%
\bibitem{Burgess:1990zu}
  M.~Burgess and D.~J.~Toms,
  Phys.\ Lett.\  B {\bf 234} (1990) 97.
%
\bibitem{Hatanaka}
 H.~Hatanaka, T.~Inami and C.~S.~Lim,
 Mod.\ Phys.\ Lett.\ A {\bf 13} (1998) 2601.
%
\bibitem{finiteHiggs}
Y.~Hosotani, N.~Maru, K.~Takenaga and T.~Yamashita,
  Prog.\ Theor.\ Phys.\  {\bf 118}, 1053 (2007); 
%
Y.~Hosotani,
  arXiv:hep-ph/0607064;  
%
\bibitem{fiveflat}
 A.~Higuchi and L.~Parker,
  Phys.\ Rev.\  D {\bf 37} (1988) 2853;  
%
M.~Kubo, C.~S.~Lim and H.~Yamashita,
  Mod.\ Phys.\ Lett.\  A {\bf 17} (2002) 2249; 
%
G.~Burdman and Y.~Nomura,
  Nucl.\ Phys.\  B {\bf 656} (2003) 3; 
%
C.~A.~Scrucca, M.~Serone and L.~Silvestrini,
  Nucl.\ Phys.\  B {\bf 669} (2003) 128; 
%
N.~Haba, M.~Harada, Y.~Hosotani and Y.~Kawamura,
  Nucl.\ Phys.\  B {\bf 657} (2003) 169  [Erratum-ibid.\  B {\bf 669} (2003) 381];
N.~Haba, Y.~Hosotani and Y.~Kawamura,
  Prog.\ Theor.\ Phys.\  {\bf 111} (2004) 265; 
N.~Haba, Y.~Hosotani, Y.~Kawamura and T.~Yamashita,
  Phys.\ Rev.\  D {\bf 70} (2004) 015010; 
N.~Haba and T.~Yamashita,
  JHEP {\bf 0404} (2004) 016; 
%
C.~S.~Lim and N.~Maru,
  Phys.\ Lett.\  B {\bf 653}, 320 (2007); 
%
\bibitem{fivewarped} 
C.~Csaki, C.~Grojean, L.~Pilo and J.~Terning,
  Phys.\ Rev.\ Lett.\  {\bf 92} (2004) 101802; 
Y.~Nomura,
  JHEP {\bf 0311} (2003) 050; 
G.~Burdman and Y.~Nomura,
  Phys.\ Rev.\  D {\bf 69} (2004) 115013; 
%
Y.~Hosotani and M.~Mabe,
  Phys.\ Lett.\  B {\bf 615} (2005) 257; 
Y.~Hosotani, K.~Oda, T.~Ohnuma and Y.~Sakamura,
  Phys.\ Rev.\  D {\bf 78}, 096002 (2008);
%
\bibitem{sixorbifold} 
C.~Csaki, C.~Grojean and H.~Murayama,
  Phys.\ Rev.\  D {\bf 67} (2003) 085012;
%
C.~A.~Scrucca, M.~Serone, L.~Silvestrini and A.~Wulzer,
  JHEP {\bf 0402} (2004) 0490 
%
Y.~Hosotani, S.~Noda and K.~Takenaga,
  Phys.\ Lett.\  B {\bf 607} (2005) 276;
%
C.~S.~Lim, N.~Maru and K.~Hasegawa,
  J.\ Phys.\ Soc.\ Jap.\  {\bf 77} (2008) 074101;
%
C.~S.~Lim and N.~Maru,
  Phys.\ Rev.\  D {\bf 75}, 115011 (2007). 
%
\bibitem{Hosotani:2004ka}
  Y.~Hosotani, S.~Noda and K.~Takenaga,
  Phys.\ Rev.\  D {\bf 69} (2004) 125014.
%
\bibitem{orbifold}
L.~J.~Dixon, J.~A.~Harvey, C.~Vafa and E.~Witten,
Nucl.\ Phys.\ B {\bf 261}, 678 (1985) and 
Nucl.\ Phys.\ B {\bf 274}, 285 (1986).
%
\bibitem{Witten:1983ux}
  E.~Witten,
  ``Fermion Quantum Numbers In Kaluza-Klein Theory,''
%
\bibitem{Rubakov}
  V.~A.~Rubakov and M.~E.~Shaposhnikov, 
  Phys.\ Lett.\ B {\bf 125}, 136 (1983);
%
  C.~G.~.~Callan and J.~A.~Harvey,
  Nucl.\ Phys.\ B {\bf 250}, 427 (1985).
%
\bibitem{Randjbar}
  S.~Randjbar-Daemi, A.~Salam, J.~Strathdee,
  Nucl.\ Phys.\ B {\bf 214}, 491 (1983).
%
\bibitem{Gross}
  D.~J.~Gross, J.~A.~Harvey, E.~J.~Martinec and R.~Rohm, 
  Phys.\ Rev.\ Lett.\  {\bf 54}, 502 (1985).
%
\bibitem{Aldazabal:2000sa}
  G.~Aldazabal, L.~E.~Ibanez, F.~Quevedo and A.~M.~Uranga,
  JHEP {\bf 0008} (2000) 002;
  G.~Aldazabal, S.~Franco, L.~E.~Ibanez, R.~Rabadan and A.~M.~Uranga,
  JHEP {\bf 0102} (2001) 047; 
  For an effective approach see also C.~P.~Burgess, D.~Hoover, C.~de Rham and G.~Tasinato,
  JHEP {\bf 0903}, 124 (2009);
  I.~Antoniadis, A.~Kumar and B.~Panda,
  Nucl.\ Phys.\  B {\bf 823} (2009) 116.  
%
\bibitem{Salvatori:2006pb}
  M.~Salvatori,
  JHEP {\bf 0706}, 014 (2007). 
%
\bibitem{Ambjorn:1980sm}
  J.~Ambjorn and H.~Flyvbjerg,
  Phys.\ Lett.\  B {\bf 97}, 241 (1980).
%
\bibitem{tHooft}
  G.~'t Hooft,
  Nucl.\ Phys.\  B {\bf 153}, 141 (1979).
 G.~'t Hooft,
  Commun.\ Math.\ Phys.\  {\bf 81} (1981) 267.
%
\bibitem{vonGersdorff:2007uz}
G.~von Gersdorff,
  Nucl.\ Phys.\  B {\bf 793}, 192 (2008); 
 G.~von Gersdorff,
  JHEP {\bf 0808}, 097 (2008).
%
\bibitem{Abe:2008fi}
 H.~Abe, T.~Kobayashi and H.~Ohki,
  JHEP {\bf 0809}, 043 (2008);
H.~Abe, K.~S.~Choi, T.~Kobayashi and H.~Ohki,
  Nucl.\ Phys.\  B {\bf 814}, 265 (2009); 
%
\bibitem{Hebecker}
  A.~Hebecker and J.~March-Russell,
  Nucl.\ Phys.\  B {\bf 625}, 128 (2002).
%
\bibitem{Quiros}
  M.~Quiros,
  arXiv:hep-ph/0302189.
%
\bibitem{SS}
  J.~Scherk and J.~H.~Schwarz,
  Nucl.\ Phys.\  B {\bf 153}, 61 (1979).
J.~Scherk and J.~H.~Schwarz,
  Phys.\ Lett.\  B {\bf 82}, 60 (1979).
%
\bibitem{Lebedev:1988wd}
  D.~R.~Lebedev, M.~I.~Polikarpov and A.~A.~Roslyi,
  Nucl.\ Phys.\  B {\bf 325} (1989) 138.
%
\bibitem{GonzalezArroyo:1982hz}
  A.~Gonzalez-Arroyo and M.~Okawa,
  Phys.\ Rev.\  D {\bf 27}, 2397 (1983).
%
\bibitem{Vassilevich:2003xt}
  D.~V.~Vassilevich,
  Phys.\ Rept.\  {\bf 388}, 279 (2003).

\bibitem{Dowker:1975tf}
  J.~S.~Dowker and R.~Critchley,
  Phys.\ Rev.\ D {\bf 13}, 3224 (1976).

\bibitem{Hawking:1976ja}
  S.~W.~Hawking,
  Commun.\ Math.\ Phys.\  {\bf 55}, 133 (1977).

\bibitem{Alvarez:2006we}
  E.~Alvarez and A.~F.~Faedo,
  JHEP {\bf 0605} (2006) 046;
  E.~Alvarez and A.~F.~Faedo,
  Phys.\ Rev.\  D {\bf 74} (2006) 124029;
  V.~P.~Frolov, P.~Sutton and A.~Zelnikov,
  Phys.\ Rev.\  D {\bf 61} (2000) 024021;
  M.~J.~Duff and D.~J.~Toms,
  ``Divergences And Anomalies In Kaluza-Klein Theories,''
  CERN-TH-3248
  {\it Presented at Second Seminar on Quantum Gravity, Moscow, USSR, Oct 13-15, 1981};
M.~J.~Duff and D.~J.~Toms,
``Kaluza-Klein Kounterterms,'' CERN-TH-3259
{\it Presented at 2nd Europhysics Study Conf. on Unification of Fundamental Interactions,
 Erice, Sicily, Oct 6-14, 1981}.
%
\bibitem{vonGersdorff:2008df}
  G.~von Gersdorff,
  JHEP {\bf 0808} (2008) 097.
%
\bibitem{Dowker:1976pr}
  J.~S.~Dowker and R.~Critchley,
  Phys.\ Rev.\  D {\bf 15} (1977) 1484;
  J.~S.~Dowker,
  J.\ Phys.\ A  {\bf 10} (1977) 115.
%
\bibitem{Giusti:2001ta}
 L.~Giusti, A.~Gonzalez-Arroyo, C.~Hoelbling, H.~Neuberger and C.~Rebbi,
 Phys.\ Rev.\  D {\bf 65} (2002) 074506;
 A.~Gonzalez-Arroyo and A.~Ramos,
 JHEP {\bf 0407} (2004) 008.


\end{thebibliography}
